\documentclass[prd,preprint,tightenlines,floatfix,showpacs,preprintnumbers,nofootinbib,eqsecnum,superscriptaddress]{revtex4-1}

\usepackage{color}      
\usepackage[dvips]{graphicx}    
\usepackage{amsbsy}     
\usepackage{amsfonts}       
\usepackage{amsmath}        
\usepackage{amssymb}        
\usepackage{color}
\usepackage{wasysym}
\usepackage{multirow}
\usepackage{mciteplus}
\usepackage{psfrag}
\usepackage{slashed}
\usepackage{cancel}
\usepackage{soul}

\usepackage{hyperref}

\newcommand{\MO}{{\tt micrOMEGAs}}
\newcommand{\SARAH}{{\tt SARAH}}
\newcommand{\FeynRules}{{\tt FeynRules}}
\newcommand{\FeynArts}{{\tt FeynArts}}
\newcommand{\FormCalc}{{\tt FormCalc}}

\newcommand{\GeV}{{\;\mathrm{GeV}}}

\newcommand{\MSbar}{{$\overline{\mathrm{MS}}$}}

\newcommand{\hn}{h^0}
\newcommand{\Hn}{H^0}
\newcommand{\An}{A^0}
\newcommand{\Hp}{H^\pm}

\newcommand{\lam}{\lambda}

\newcommand{\Mlop}{M_{\mathrm{LOP}}}
\newcommand{\lamL}{\lambda_L}
\newcommand{\lamS}{\lambda_S}

\newcommand{\lm}{\overline{m}}

\catcode`\@=11
\font\manfnt=manfnt
\def\Watchout{\@ifnextchar [{\W@tchout}{\W@tchout[1]}}
\def\W@tchout[#1]{{\manfnt\@tempcnta#1\relax%
  \@whilenum\@tempcnta>\z@\do{%
    \char"7F\hskip 0.3em\advance\@tempcnta\m@ne}}}
\let\foo\W@tchout
\def\dubious{\@ifnextchar[{\@dubious}{\@dubious[1]}}

\def\@dubious[#1]{%
  \setbox\@tempboxa\hbox{\@W@tchout#1}
  \@tempdima\wd\@tempboxa
  \list{}{\leftmargin\@tempdima}\item[\hbox to 0pt{\htr{\hss\@W@tchout#1}}]}
\def\@W@tchout#1{\W@tchout[#1]}
\catcode`\@=12

\begin{document}

\begin{flushright}
	LAPTH-006/13
\end{flushright}
\vspace{1cm}

\title{Dark matter in the Inert Doublet Model\\ after the discovery of a Higgs-like boson at the LHC}
\vspace{5em}

\author{A.~Goudelis}
 \email{andreas.goudelis@lapth.cnrs.fr}
 \affiliation{
  	LAPTh, Universit\'e de Savoie, CNRS, 9 Chemin de Bellevue, B.P.\ 110, F-74941 Annecy-le-Vieux, France
 	     }

\author{B.~Herrmann}
 \email{herrmann@lapth.cnrs.fr}
 \affiliation{
	LAPTh, Universit\'e de Savoie, CNRS, 9 Chemin de Bellevue, B.P.\ 110, F-74941 Annecy-le-Vieux, France
	     }

\author{O.~St{\aa}l}
 \email{oscar.stal@fysik.su.se}
 \affiliation{
 	The Oskar Klein Centre, Department of Physics\\
	Stockholm University, AlbaNova, SE-106 91 Stockholm, Sweden
 	     }


\begin{abstract}
We examine the Inert Doublet Model in light of the discovery of a
Higgs-like boson with a mass of roughly $126$~GeV at the LHC. We evaluate one-loop corrections to the scalar masses and perform a numerical solution of the one-loop renormalization group equations. Demanding vacuum stability, perturbativity, and $S$-matrix unitarity, we compute the scale up to which
the model can be extrapolated. From this we derive constraints on the model parameters in the presence of a 126~GeV Higgs boson. We perform an improved calculation of the dark matter relic density with the Higgs mass fixed to the measured value, taking into account the effects of three- and four-body final states resulting from off-shell production of gauge bosons in dark matter annihilation. Issues related to direct detection of dark matter are discussed, in particular the role of hadronic uncertainties. The predictions for the interesting decay mode $h^0 \rightarrow \gamma\gamma$ are presented for scenarios which fulfill all model constraints, and we discuss how a potential enhancement of this rate from the charged inert scalar is related to the properties of dark matter in this model. We also apply LHC limits on Higgs boson decays to invisible final states, which provide additional constraints on the mass of the dark matter candidate. Finally, we propose three benchmark points that capture different aspects of the relevant phenomenology.
\end{abstract}


\maketitle

\newpage


\section{Introduction}

The Inert Doublet Model (IDM) is the simplest among the models with two Higgs doublets. In addition to the Standard Model (SM) 
particle content, it contains an extra doublet of complex scalar fields which couples to the SM scalar and gauge boson sector but not to the 
fermions. Moreover, it involves a discrete $\mathbb{Z}_2$ symmetry under which the new scalar doublet is odd and all the other particles are 
even, which makes that the new ``inert'' doublet particles can only appear in even number in interaction vertices.

The IDM was first introduced more than three decades ago in studies of electroweak symmetry breaking (EWSB) \cite{Deshpande:1977rw}. 
Long after, it was proposed as a model that can provide a viable dark matter candidate according to the thermal relic picture 
\cite{Ma:2006km, Barbieri:2006dq}, since the neutral scalars contained in the new doublet can be seen as weakly interacting massive 
particles (WIMPs) and play the role of the dark matter (DM) in our universe. Due to its rich phenomenology for cosmology and particle 
physics, the IDM has received considerable attention \cite{Gustafsson:2007pc, *Agrawal:2008xz, *Andreas:2009hj, *Nezri:2009jd, *Arina:2009um, *Gong:2012ri}. 
Its DM candidate captures all the basic mechanisms through which the observed relic density can be generated in WIMP models 
\cite{LopezHonorez:2006gr}: The ``correct'' relic abundance \cite{Komatsu:2010fb} can be achieved by adjusting couplings, by approaching or taking distance from resonances, or by co-annihilating with another particle.\footnote{Coannihilation is absent, e.g., in the 
simpler singlet scalar model.}

Additionally, the IDM was advocated to allow for a heavier SM-like Higgs boson compatible with electroweak precision tests, 
$m_{\hn}\gtrsim 200\GeV$, without resorting to unnatural fine-tuning \cite{Barbieri:2006dq} (for recent work also considering 
this possibility, see, e.g., Ref.\ \cite{Gustafsson:2012aj}). The new states predicted by the IDM have been subjected to collider 
bounds \cite{Lundstrom:2008ai}, and they provide an interesting phenomenology for the Large Hadron Collider (LHC) 
\cite{Barbieri:2006dq, Cao:2007rm, *Dolle:2009ft, *Miao:2010rg, *Wang:2012zv, *Osland:2013sla}. While by now a heavy Higgs boson with SM couplings is experimentally ruled out \cite{ATLASDiscovery, CMSDiscovery}, the model still remains attractive due to its dark matter features. Furthermore, the IDM
provides an interesting example of interplay between DM and Higgs physics, since the SM-like Higgs boson is one of the basic means of 
communication between the ``dark''\footnote{This term is used in a slightly abusive way here, since the model does not contain what is 
usually dubbed a ``dark sector'' in the litterature.} and ``visible'' sectors of the model. Hence, both the mass and the couplings of 
the SM-like Higgs boson are of crucial importance to assess whether the IDM can indeed explain the dark matter in the universe. 

With the recent announcement of the observation of a Higgs-like resonance with a mass of around 125--126 GeV by the ATLAS and CMS 
collaborations \cite{ATLASDiscovery, CMSDiscovery}, as well as the supporting hints from the D{\O} and CDF experiments \cite{TCB:2012zzl}, 
it appears likely that the particle responsible for EWSB (or at least one of them) has been discovered. In this spirit, we find it timely 
and interesting to examine implications of this observation for the IDM. Apart from the obvious consequence that the number of free 
model parameters is reduced by one, other interesting features appear, as we shall describe in the present paper. Moreover, despite the attention that the IDM has received in the community, most studies rely on lowest order predictions only. An exception is the study beyond leading order that was performed in Ref.\ \cite{Hambye:2007vf}. This work is, however, limited to the interesting case of radiative electroweak symmetry breaking \`a la Coleman-Weinberg, a scenario leading to rather extreme parameter values. Another exception is the very recent paper \cite{Klasen2013EWDD}, where higher-order corrections to DM direct detection in the IDM are calculated. 

In the present work, we perform an analysis of the IDM parameter space assuming that the LHC is indeed observing a (SM-like) Higgs boson with a mass $M_{\hn} \approx 126$ GeV. In Sec.\ \ref{Sec:TheModel}, we introduce the model and the relevant notation. In Sec.\ \ref{Sec:IDMoneloop} we present one-loop corrections to the scalar masses in the IDM, and expressions for the one-loop renormalization group equations (RGEs) for the model's quartic couplings, which are used  to study vacuum stability, perturbativity, and unitarity constraints. Experimental constraints from collider, low-energy, and cosmological data are presented in Sec.\ \ref{Sec:ConstraintsGeneral}. The corresponding numerical analysis, which contains the main results of this work, is presented in Sec.\ \ref{Sec:Analysis}, where we perform extensive scans over the model parameter space. The implications for dark matter and the interesting Higgs decay into two photons are discussed in detail. Based on our analysis, we identify benchmark scenarios capturing general features of the parameter space. Finally, Sec.\ \ref{Sec:Discussion} contains a summary of our results and the main conclusions.

\section{The Inert Doublet Model at tree level \label{IDMgeneral} \label{Sec:TheModel}}
The inert doublet model (IDM) contains, in addition to the Standard Model (SM) particle content, a second complex scalar doublet. The model Lagrangian is constructed so as to satisfy an exact $\mathbb{Z}_2$ symmetry (parity) under which all SM particles, including one of the scalar doublets, are even and the second scalar doublet is odd. Since this discrete symmetry prevents mixing between the scalars, one of the doublets, denoted by $H$, is similar to the SM Higgs doublet. Given that the second doublet, which we denote by $\Phi$, has odd $\mathbb{Z}_2$ parity it is inert in the sense that its component fields do not couple singly to SM particles. The requirement of renormalizability then also forbids all tree-level couplings to the fermion sector. 

Imposing the $\mathbb{Z}_2$ symmetry has two further important consequences. First, the inert doublet $\Phi$ does not acquire a vacuum expectation value. Second, it forbids several of the terms appearing in the general two-Higgs-doublet model scalar potential \cite{Branco:2011iw}. More precisely, the tree-level scalar potential of the IDM takes the form
\begin{equation}
	V_0 ~=~ \mu_1^2 |H|^2  + \mu_2^2|\Phi|^2 + \lambda_1 |H|^4+ \lambda_2 |\Phi|^4 + \lambda_3 |H|^2| \Phi|^2
		+ \lambda_4 |H^\dagger\Phi|^2 + \frac{\lambda_5}{2} \Bigl[ (H^\dagger\Phi)^2 + \mathrm{h.c.} \Bigr].
\label{Eq:TreePotential}
\end{equation}
In the general case, the $\lambda_i$ are complex parameters. Although considering this possibility can have interesting consequences for CP-violation and electroweak baryogenesis \cite{Chowdhury:2011ga, Borah:2012pu, Cline:2013bln}, in this work we limit ourselves to the case of real values. Upon electroweak symmetry breaking, the two doublets can be expanded in components as
\begin{equation}
	H ~=~ \left( \begin{array}{c} G^+ \\ \frac{1}{\sqrt{2}}\left(v+\hn+\mathrm{i}G^0\right) \end{array} \right),
	\qquad
	\Phi ~=~ \left( \begin{array}{c} H^+\\ \frac{1}{\sqrt{2}}\left(H^0+\mathrm{i}A^0\right) \end{array} \right),
\end{equation}
where $v = \sqrt{2}~\langle 0 | H | 0 \rangle \approx 246$ GeV denotes the vacuum expectation value of the 
neutral component of the doublet $H$. The $\hn$ state corresponds to the physical SM-like Higgs-boson, whereas 
$G^0$ and $G^{\pm}$ are the Goldstone bosons. The inert sector consists of a neutral CP-even scalar $\Hn$, a pseudo-scalar $\An$, and a pair of charged scalars $\Hp$.

A phenomenologically important consequence of the $\mathbb{Z}_2$ symmetry is that the lightest $\mathbb{Z}_2$-odd particle (LOP) is stable. If further the LOP is either $\Hn$ or $\An$, this (neutral) state can play the role of the DM candidate, analogously to the Lightest Supersymmetric Particle (LSP) in supersymmetric models with $R$-parity conservation.

At the tree level, the scalar masses are obtained from the potential in Eq.\ \eqref{Eq:TreePotential}. When the potential is expanded in the component fields, the masses of the physical states are given by
\begin{align}
	m_{\hn}^2 &= \mu_1^2 + 3 \lambda_1 v^2, \\ 	
	m_{\Hn}^2 &= \mu_2^2 + \lambda_L v^2, \label{Eq:mH0tree} \\
	m_{\An}^2 &= \mu_2^2 + \lambda_S v^2, \\
	m_{\Hp}^2 &= \mu_2^2 + \frac{1}{2} \lambda_3 v^2.
\end{align}
Here we have introduced the useful abbreviations
\begin{eqnarray}
	\lambda_L &=& \frac{1}{2} \left( \lambda_3 + \lambda_4 + \lambda_5 \right), \\
	\lambda_S &=& \frac{1}{2} \left( \lambda_3 + \lambda_4 - \lambda_5 \right).
\end{eqnarray}
As in the Standard Model, we have the scalar potential minimization relation $m_{\hn}^2 = -2\mu_1^2 = 2 \lam_1 v^2$ that can be used to eliminate the parameter $\mu_1^2$ after electroweak symmetry breaking.

The IDM scalar sector can hence be specified by a total of six parameters
\begin{equation}
\left\{ \lam_1, ~~ \lam_2, ~~ \lam_3, ~~ \lam_4, ~~ \lam_5, ~~ \mu_2 \right\},
\label{eq:lambdas}
\end{equation}
which can be exchanged through the above equations in favour of the physically more meaningful set 
\begin{equation}
 	\left\{ m_{\hn}, ~~ m_{\Hn}, ~~ m_{\An}, ~~ m_{\Hp}, ~~ \lambda_L, ~~ \lambda_2 \right\},
	\label{eq:masses}
\end{equation} 
that is often used in phenomenological applications. It is worth noting that the parameter $\lamL$ has a simple physical interpretation as the $\hn-\Hn-\Hn$ coupling at the tree level (similarly, $\lamS$ is the $\hn-\An-\An$ coupling). In contrast, the parameter $\lambda_2$, which is common to both parameter sets, appears only in quartic self couplings among inert particles and does therefore not enter any physically observable process \textit{at the tree level}. However, it plays a role once the theory is considered beyond leading order. In the following, we shall make use of both sets of parameters given here. 

\section{The Inert Doublet Model beyond the tree level \label{Sec:IDMoneloop}}

Most of the work that has been performed within the IDM so far, such as collider signals, vacuum stability considerations or relic abundance constraints, has been based on pure leading-order calculations.
Generally speaking, it is meaningful to go beyond tree level when theoretical uncertainties of some calculation start to become comparable to the corresponding experimental uncertainties. In many models this is the case, for example, for the well-measured relic abundance of dark matter in our universe. Another reason can be to investigate effects which are absent at tree level but arise in leading order at the one-loop level as it is the case, e.g., for the decay $h^0 \rightarrow \gamma\gamma$. 

In the IDM, radiative corrections to inert scalar annihilations, which are relevant for DM phenomenology, are expected to be rather small \cite{Drees:2009gt}. Here, strong couplings do not intervene in the relevant Feynman diagrams (with the exception of a final state vertex correction involving gluon exchange among final state quarks), contrary to the Minimal Supersymmetric Standard Model, where ${\cal O}(\alpha_s)$ corrections can have a sizeable impact in this context \cite{Freitas:2007sa, *Herrmann:2007ku, *Herrmann:2009wk, *Herrmann:2009mp, *Harz:2012fz}. Moreover, the number of diagrams with electroweak couplings at one-loop order is rather limited, again unlike the case of the MSSM, where electroweak corrections can become significant due to the very large number of contributing diagrams \cite{Baro:2007em, *Baro:2009na, *Boudjema:2011ig}.

With the electroweak scale being probed by both colliders and dark matter detection experiments, the different new physics models are constrained and the allowed magnitude of their couplings becomes limited. In particular, the IDM parameter space is known to present three mass regimes for the lightest odd particle (LOP), where the observed relic density can be obtained \cite{Barbieri:2006dq, LopezHonorez:2006gr, Hambye:2009pw, Honorez:2010re, LopezHonorez:2010tb}: the so-called low- ($1\GeV \lesssim M_{\rm LOP} \lesssim 10\GeV$), intermediate- ($10\GeV \lesssim M_{\rm LOP} \lesssim 150\GeV$), and high-mass ($M_{\rm LOP} \gtrsim 500\GeV$) regimes (the distinction being somewhat arbitrary). As we shall see in the following, the low- and intermediate-mass regimes are already quasi-excluded from the XENON100 experiment results \cite{Aprile:2012nq}. As a consequence, in order for the IDM to still provide DM-compliant scenarios in this mass range, one has to rely on resonances and thresholds to achieve
the correct relic density without violating the null XENON results, which would happen for sufficiently large
coupling values. In particular, we will see that the only points of the interesting intermediate mass regime that survive the
constraint are lying close to the Higgs ``funnel'' region and the $WW^*$ production threshold (see also \cite{Gustafsson:2012aj, Klasen2013EWDD} for recent works on the subject), i.e.\ the region where LOP annihilation is dominated by nearly on-shell $\hn$ exchange or gauge boson pair-production. 

In this regime, given the accuracy in the experimental determination of the relic density, a mass shift 
of the order of $2$ GeV is sufficient to render a particular point in parameter space excluded or not.\footnote{Note that this mass difference is also similar to the current difference between the values for the Higgs mass extracted by ATLAS and CMS in different channels.} If radiative corrections are sufficiently large to shift the LOP mass for which the correct relic density is observed in or out of the viable region, then the computation of higher-order corrections can become important. This is well-known in the case of supersymmetric models such as the MSSM or the NMSSM, see, e.g., Ref.\ \cite{Staub:2010ty}. The exact values of the involved masses are also known to be essential in the case of co-annihilation of the LOP with the next-to-lightest $\mathbb{Z}_2$-odd particle (NLOP). In this case, a shift in one or both masses can change the importance of this particular channel and thus the prediction of the relic abundance.

In order to perform arbitrary calculations at the one-loop level within the IDM, we have implemented the 
particle content and the interactions (in 't Hooft-Feynman gauge) in model files for the Mathematica package \FeynArts/\FormCalc~\cite{Hahn:2000kx}. The implementation has been performed in parallel and cross-checked with the help of the two packages \FeynRules~\cite{Christensen:2008py} and \SARAH~\cite{Staub:2008uz,*Staub:2009bi,*Staub:2010jh}.

\subsection{One-loop corrections to the scalar masses \label{Sec:DiagrammaticCalculation}}

In the Feynman-diagrammatic approach, the one-loop corrected scalar masses are obtained by evaluating the  corrections to the two-point functions. Denoting the mass for a generic scalar field $\phi$ appearing in the renormalized Lagrangian by $m_\phi$, the corresponding (renormalized) inverse propagator, $\Gamma_\phi$, can be written as
\begin{equation}
	\Gamma_\phi(p^2) = \mathrm{i}\left[p^2-\left(m_\phi^2 - \mathrm{Re}\,\Sigma_\phi(p^2)\right)\right],
\label{eq:twopoint}
\end{equation}
where $\Sigma_\phi(p^2)$ is the (again, renormalized) self-energy of the field $\phi$. From this equation, two further concepts can be defined. The first is that of the \emph{running mass}, which we denote by $\overline{m}_\phi(Q)$. As indicated by the argument, this mass carries an explicit dependence on the renormalization scale $Q$. The running mass is obtained from the inverse two-point function evaluated at zero external momentum\footnote{An alternative method which yields the same expression for the running masses is through the (one-loop) effective potential, see \cite{Hambye:2007vf} for the case of the IDM.} 
\begin{equation}
	\lm_\phi^2(Q) = \mathrm{i}\Gamma_\phi(p^2=0)=m_\phi^2-\mathrm{Re}\,\Sigma_\phi(p^2=0).
\label{eq:runm}
\end{equation}
The second type of one-loop mass is obtained by solving Eq.~\eqref{eq:twopoint} for the value of $p^2=M_\phi^2$ that gives $\Gamma(M_\phi^2)=0$. In practice, this equation is solved iteratively with usually very fast convergence.
Since it corresponds to a pole of the propagator, $M_\phi$ is refered to as the \emph{pole mass}.
Which one-loop mass definition should be used depends on what is most convenient in a given situation. Working in fixed order of the perturbative expansion, the impact of this choice on physical observables is a higher-order effect. Below we shall make use of both the running mass and the pole mass when studying different aspects of the IDM.
\begin{figure}
	\includegraphics[scale=0.8]{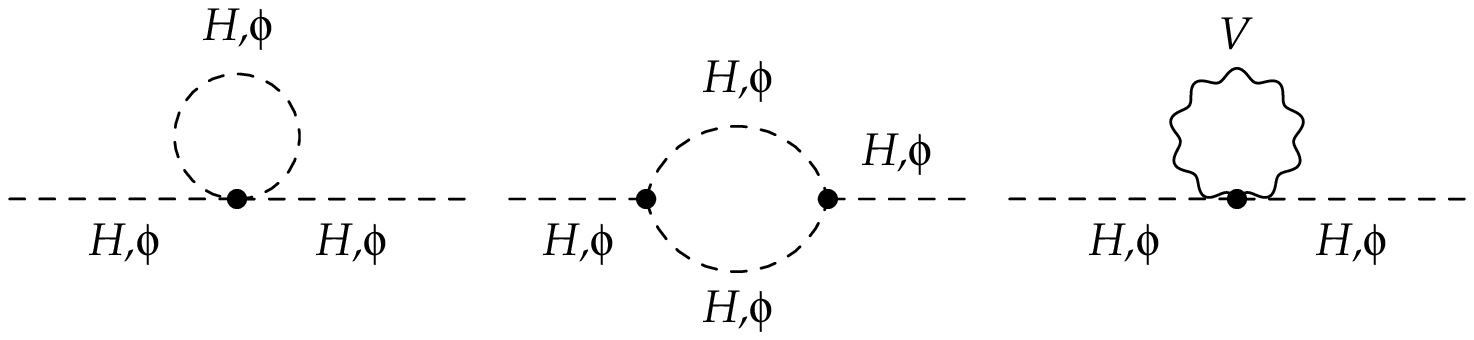}\\
	\includegraphics[scale=0.53333]{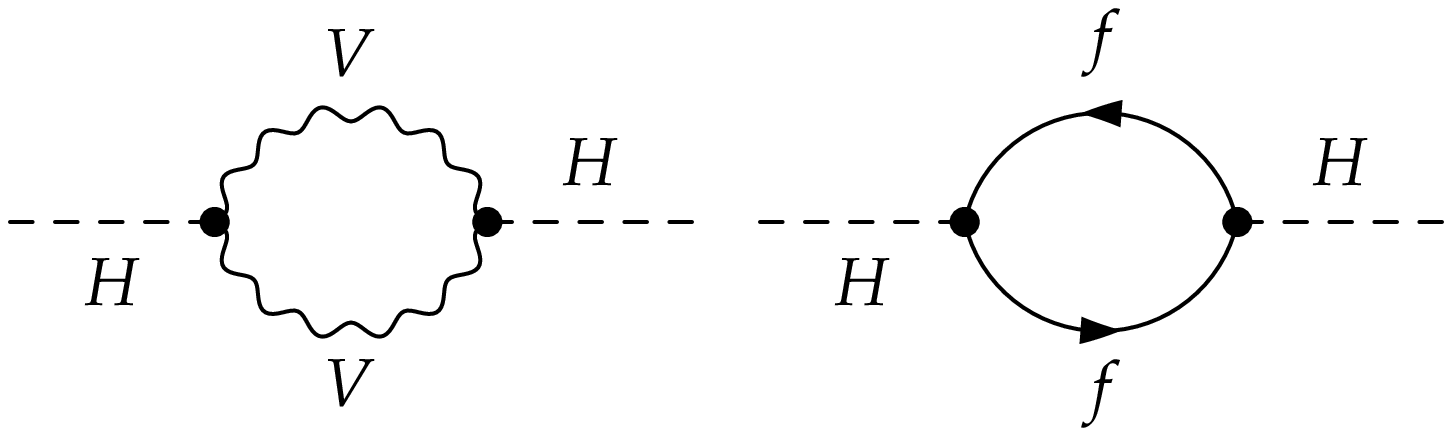}
	\caption{Generic classes of Feynman diagrams contributing to the one-loop self-energy of scalars in the 
Inert Doublet Model. The upper diagrams involving scalars and vector bosons ($V$) contribute to the masses of both the SM-like ($H$) and the inert scalar doublet ($\Phi$), while the lower diagrams involving a vector boson ($V$) or a fermion ($f$) are only present for the SM-like Higgs doublet.}
	\label{Fig:SelfEnergy}
\end{figure}

The one-loop self-energies appearing in Eq.~\eqref{eq:twopoint} generally receive contributions from scalars, gauge bosons, and fermions. In the IDM, the diagrams involving fermions are restricted to the SM-like Higgs boson, which is also the only state that couples singly to the vector bosons. Generic classes of diagrams contributing to the scalar self-energies are shown in Fig.\ \ref{Fig:SelfEnergy}. Working in the \MSbar\ renormalization scheme, we evaluate these in the 't Hooft-Feynman gauge, which leads to the expressions for the running masses given in Appendix \ref{App:ScalarMasses}.
Before showing numerical results, we should also mention that these results have been checked against those presented in Ref.\ \cite{Hambye:2007vf}, and overall agreement was found (modulo the fact that Ref.\ \cite{Hambye:2007vf} does not include the $\mathcal{O}(\alpha)$ corrections coming from vector boson loops). Moreover, the SM-type contributions for $\lm_{\hn}$ have been checked against \cite{Denner:1991kt}.

\begin{figure}
	\begin{center}
		\includegraphics[width=0.95\columnwidth]{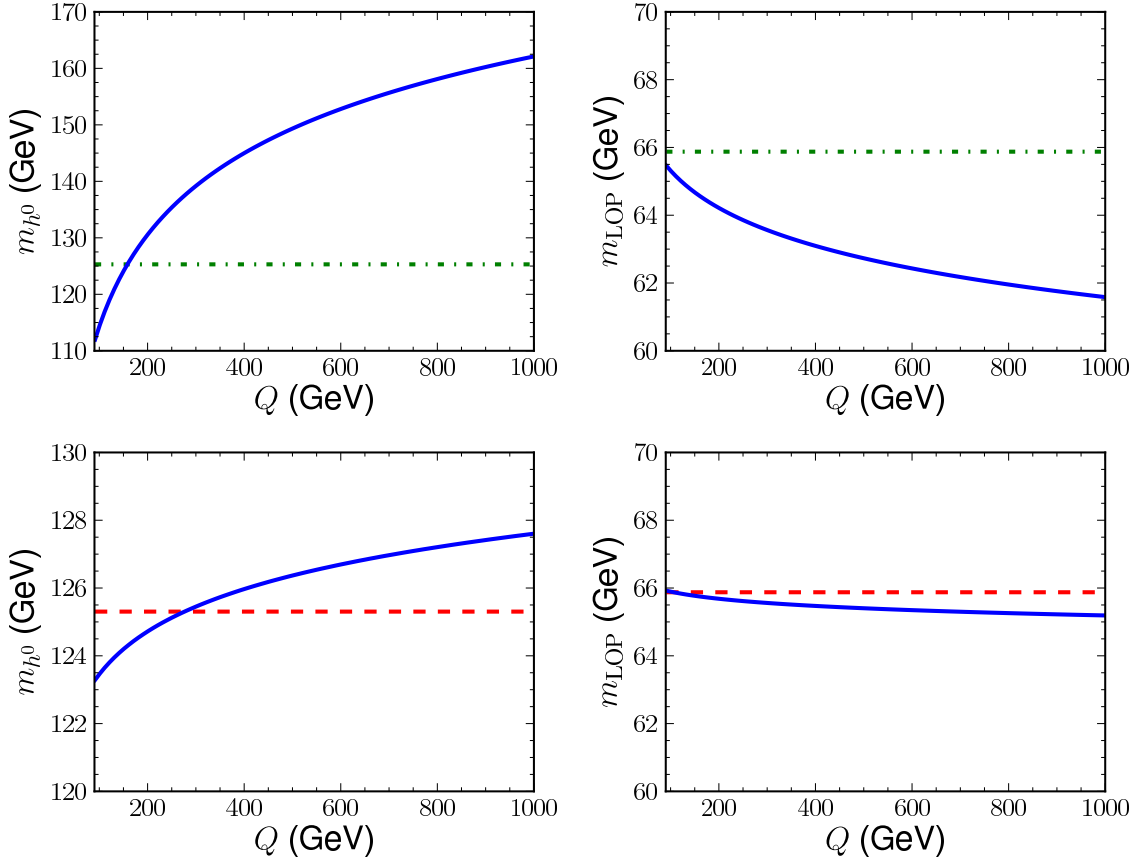}
	\end{center}
	\vspace{-2em}
	\caption{Running one-loop masses in the \MSbar\ scheme of the SM Higgs boson ($\hn$) and the lightest inert scalar (LOP) as a function of the renormalization scale $Q$ (blue, solid lines). This example corresponds to the input parameters $m_{\hn}=125.3$ GeV, $m_{\Hn}=66.0$ GeV, $m_{\An}=131.0$ GeV, $m_{\Hp}=137.0$ GeV, $\lamL=0.12$, and $\lam_2=8.6\cdot 10^{-4}$. In the upper panels, the running mass is obtained by interpreting the input mass as the tree-level mass (green, dash-dotted), while in the lower panel the input value is interpreted as the pole mass (red, dashed). In this example, the LOP is the CP-even scalar $\Hn$.}
	\label{Fig:Masses1}
\end{figure}

We find that the relative difference between the \MSbar\ and the tree-level masses can, depending on the exact parameter values and the chosen renormalization scale, reach up to 30\% for the SM-like Higgs boson $\hn$, and up to around 10\% for the inert scalars. In Fig.\ \ref{Fig:Masses1}, this is illustrated for an example scenario.\footnote{This scenario corresponds to the benchmark point I defined in Tab.\ \ref{tab:bench} below.}
As expected, the difference between the pole and the running \MSbar\ masses is smaller. For the example in Fig.~\ref{Fig:Masses1}, it amounts to up to 2 GeV (about 1.5\%) for $\hn$ and up to almost 1~GeV (about 1.5\%) for $\Hn$, which is representative for typical parameter points in the IDM. The situation is similar for the masses of the other inert scalars, $\An$ and $\Hp$, which are not shown here.

\subsection{Renormalization group equations for the quartic couplings}

Renormalization group equations (RGEs), when evaluated at one-loop accuracy, resum leading scale logarithms to all orders. They are therefore useful for studying the evolution of parameters under renormalization over a large energy range. To treat all parameters at a similar footing, we find it more convenient here to work in the basis of quartic couplings, Eq.~\eqref{eq:lambdas}, rather than the mass basis. The renormalization group equations for the quartic couplings $\lambda_i$ $(i=1\ldots 5)$ are driven by the corresponding beta functions, $\beta_{\lambda_i}$, according to
\begin{equation}
	16\pi^2\frac{\partial \lambda_i}{\partial \log Q} ~=~ \beta_{\lambda_i}
	~=~ \beta^{(s)}_{\lambda_{i}}+\beta^{(g)}_{\lambda_{i}}+\beta^{(y)}_{\lambda_{i}}.
\label{eq:rge}
\end{equation}
The latter receive contributions from scalars ($s$), gauge bosons ($g$), and fermions ($y$). The different pieces of $\beta_{\lambda_i}$ are obtained by diagrammatic calculation.\footnote{The renormalization group equations can also be computed by demanding scale invariance of the one-loop effective potential. For an example of this approach for the singlet scalar model, see Ref.\ \cite{Gonderinger:2009jp}.} The results for the different contributions are given in Appendix~\ref{App:BetaFunctions}. Our expressions are in agreement with earlier results for the general 2HDM \cite{Inoue:1979nn, Ferreira:2009jb} (recalling that in the IDM $\lambda_6 = \lambda_7 = 0$).

\section{Constraints \label{Sec:ConstraintsGeneral}}

The various constraints which can be imposed on the (tree-level) parameters of the inert doublet model (IDM) have been extensively discussed in the literature. A summary is, e.g., given in \cite{Gustafsson:2010zz, *Swiezewska:2012ej}. For completeness, we dedicate this Section to discuss which constraints are relevant here and how they are treated in our analysis.

\subsection{Theoretical constraints}

A first class of constraints on the inert doublet model (IDM) comes from the minimal requirement for a physically sensible and (perturbatively) calculable theory. First, it is required that the couplings should not be larger than some value which makes a perturbative treatment meaningless. We thus demand all quartic scalar, gauge, and Yukawa couplings to fulfill a common constraint
\begin{equation}
	|\lam_i|, ~ |g_i|, ~ |y_i| ~\leq~ K .
	\label{eq:pert}
\end{equation}
A minimal condition is that this inequality be respected at the input scale where the parameters on the left-hand side are specified. The value of $K$ which should be imposed for a valid perturbative expansion is then somewhat process-dependent. Considering the more stringent requirement that Eq.~\eqref{eq:pert} remains valid under renormalization group evolution, we can use Eq.~\eqref{eq:rge} for guidance. Since the coefficients appearing in the beta functions for the scalar quartic couplings are all $\mathcal{O}(1)$, it is clear that an instability (Landau pole) will appear for $|\lambda_i(Q)|>4\pi$ in the absence of accidental cancellations. In practice, smaller values of $|\lambda_i|$ can also lead to instabilities, but we choose to be conservative and impose Eq.~(\ref{eq:rge}) with $K=4\pi$ as the perturbativity limit at all scales.

A second requirement is that the scalar potential, Eq.~\eqref{Eq:TreePotential}, should be bounded from below and that upon electroweak symmetry breaking (EWSB) it develops a minimum which renders the electroweak vacuum stable, or metastable with a sufficient long lifetime. The criterion we shall discuss here implies \emph{absolute} stability of the symmetry-breaking vacuum, which leads to conservative limits on the IDM parameters. Allowing also metastable configurations would result in relaxed constraints; we leave a detailed analysis of this question for future work.\footnote{See~Ref.\ \cite{EliasMiro:2011aa} for a recent discussion of stability/metastability in the SM.} For absolute vacuum stability to hold we demand \cite{Sher:1988mj,Branco:2011iw} 
\begin{equation}
\begin{aligned}
	\lam_1(Q), \lam_2(Q) &> 0 , \\
	\lam_3(Q) &> -2\sqrt{\lam_1(Q) \lam_2(Q)} , \\
	\lam_3(Q)+\lam_4(Q)-|\lam_5(Q)| &> -2\sqrt{\lam_1(Q) \lam_2(Q)}
	\label{eq:stability}
	\end{aligned}
\end{equation}
at the scale $Q$. 

A final constraint comes from the requirement that the scattering matrix ($S$-matrix) of every quantum field theory must be unitary. In the case of weakly coupled theories, it is sensible to require that the tree-level scattering matrix elements satisfy unitarity limits, which corresponds to imposing upper bounds on them. For the general 2HDM, the bounds were first derived in \cite{Huffel:1980sk, *Maalampi:1991fb}. Here, we use the form for the eigenvalues of the scalar and vector scattering matrices of \cite{Ginzburg:2005dt}, and we require them to be smaller than $16 \pi$. This corresponds to saturation of the unitarity limit with the tree-level contribution.

\subsection{Oblique parameters}
\label{Sec:oblique}
In models where the dominant effects of new physics appear as corrections to self energies of the (SM) gauge bosons, the effects can be parametrized in terms of the three ``oblique'' (Peskin-Takeuchi) parameters $S$, $T$, and $U$ \cite{Altarelli:1990zd, *Peskin:1991sw}, which vanish in a pure SM calculation. In the 2HDM, the contributions to the $U$ parameter are negligible, which makes it convenient to work in the approximation $U=0$. This assumption has been verified explicitly in our numerical analysis. For the case of $U=0$, recent experimental limits on the remaining two parameters are \cite{Baak:2011ze}
\begin{equation}
	S = 0.06 \pm 0.09,\quad T = 0.10 \pm 0.08. 
	\label{eq:STU}
\end{equation}
These values are based on a reference (SM) Higgs mass of $m_h^{\rm ref} = 120$ GeV and a reference top mass of $m_t^{\rm ref} = 173$ GeV.  We impose the limits resulting from Eq.~\eqref{eq:STU} at the $2\,\sigma$ confidence level as constraints on the IDM contribution. 

\subsection{Collider searches}
\label{Sec:collider}
The first constraint from direct searches at colliders comes from the invisible decay width of the $Z$ boson. If the decay mode $Z \to \Hn \An$ is open, the 
subsequent decay $\An \to \Hn f \bar{f}$ (or $\Hn \to \An f \bar{f}$ for the inverse mass hierarchy) would lead to  $Z$ decay events with fermion-antifermion pairs ($f \bar{f}$) and missing energy in the final state. A detailed analysis has shown that this decay is incompatible with LEP data, which implies that the decay width of $Z \to \Hn \An$ must be small. It is convenient to implement this constraint as \cite{Gustafsson:2007pc, Cao:2007rm}
\begin{equation}
	M_{\Hn} + M_{\An} ~\gtrsim~ M_Z.
\end{equation}
Assuming a fixed mass hierarchy $M_{\Hn} < M_{\An}$, a more detailed analysis of the IDM parameter space with respect to LEP data leads to the limit $M_{\An} ~\gtrsim~ 100~{\rm GeV}$ \cite{Lundstrom:2008ai}. 
Considering both possible mass hierarchies between $\Hn$ and $\An$, we require
\begin{equation}
	{\rm max}\left\{M_{\Hn}, M_{\An} \right\} ~\gtrsim~ 100~{\rm GeV} .
\end{equation}
Finally, limits on the mass of the charged scalar can be obtained by considering their potential pair production and subsequent decay into neutral 
Higgs bosons at LEP. Converting existing limits on the search for charginos and neutralinos, which present the same final state topology 
at colliders, leads to the bound $M_{\Hp} \gtrsim 70 - 90 \ \mathrm{GeV}$ \cite{Pierce:2007ut}.
For practical reasons, we adopt the intermediate limit
\begin{equation}
	M_{\Hp} ~\gtrsim~ M_W.
\end{equation}
In order to have a neutral DM candidate, we include as a final requirement
\begin{equation}
	M_{\Hp} > \Mlop = \min \left\{ M_{\Hn},M_{\An} \right\}
\end{equation}
for defining viable points in parameter space throughout our analysis.

\subsection{Dark matter relic density}

The recent results from the WMAP satellite, combined with other cosmological measurements, constrain the dark matter relic density to $\Omega_{\rm CDM}h^2 ~=~ 0.1126 \pm 0.0036$ \cite{Komatsu:2010fb}.
In most of the numerical analysis below, we require the IDM relic density (considering either $\Hn$ and $\An$ as the dark matter candidate) to respect this limit within $3\,\sigma$, i.e.\ we demand $\Omega_{\rm LOP}h^2$ to lie within the interval
\begin{equation}
	0.1018 \le \Omega_{\rm LOP}h^2 \le 0.1234 .
	\label{Eq:RelicDensity}
\end{equation}
In some cases we shall however relax this requirement and only impose the upper bound,
\begin{equation}
  \Omega_{\rm LOP}h^2 \le 0.1234,
	\label{Eq:RelicDensityUpper}
\end{equation}
which corresponds to the situation where the IDM contribution is only partly responsible for the observed DM density.

A subtle point in the calculation of the relic density concerns the intermediate mass regime, and in particular the mass interval $50$--$80$ GeV, where the IDM is known to produce values for $  \Omega_{\rm LOP}h^2$ in the correct range. It has been pointed out \cite{Honorez:2010re, LopezHonorez:2010tb} that in this mass regime, which lies close to the $WW$ final state threshold, contributions to the total self-/co-annihilation cross section coming from three-body final states can be substantial or even dominant. In particular the $WW^*$ contribution (and to a lesser extent also $ZZ^*$) should be taken into account to obtain reliable predictions. For this purpose we use a modified version of \MO\ \cite{Belanger:2006is, *Belanger:2008sj} which includes three-body final states' contributions. This \MO\ version also accounts for four-body final states with two virtual gauge bosons. We include these contributions in our analysis, and we find that they are not always negligible; the effect on the computed relic density can be up to a few percent.\footnote{We have noted small differences with respect to the results presented in Ref.~\cite{Honorez:2010re}, which we attribute to the inclusion of the doubly virtual final states.} For more details on the implementation of the relic density calculation we refer to the corresponding manual \cite{MOtoappear}.

\subsection{Dark matter direct detection}
\label{sect:direct}
Further constraints stem from (null) direct searches for dark matter. Until recently, the most stringent bounds on the spin-independent WIMP-nucleon scattering cross-section came from the XENON10 \cite{Angle:2011th} and XENON100
\cite{Aprile:2011hi} experiments for the low- and intermediate-mass regimes respectively. Recently, the XENON100 experiment presented updated constraints on the spin-independent cross-section, which strengthen the previous bounds by a factor of roughly $2$--$6$ \cite{Aprile:2012nq}. We impose these more stringent bounds in the numerical analysis that follows.

Our calculation of the relevant spin-independent WIMP-nucleon cross sections is again performed using the \MO\ package.
It is well-known that predictions of direct detection rates are not without uncertainties. One among the most important sources of uncertainty (at least from the particle physics side) is related to the strange quark form factor of the nucleon, $f_{Ts}$, which loosely speaking describes the ``strange quark content'' of the nucleon. This quantity is usually parametrized in terms of the so-called pion-nucleon sigma term $\sigma_{\pi N}$ \cite{Ellis:2008hf}. The value
of $\sigma_{\pi N}$ relies on experimental data that is at present poorly known. The values which are widely used come mostly from lattice QCD and chiral perturbation theory calculations \cite{Ellis:2008hf, Ohki:2008ff, *Cao:2010ph, *Alarcon:2011zs}.

Hadronic form factors, and notably $f_{Ts}$, are of particular importance when the 
dominant mechanism for WIMP-nucleon scattering is Higgs exchange (as is the case in the IDM), since the Higgs 
boson couples preferentially to heavy quarks and the strange-quark density is sufficiently large to provide a non-negligible contribution to the scattering cross section. Since the results of varying $f_{Ts}$ can be quite dramatic 
\cite{Ellis:2008hf, das:2010kb}, we perform our calculations adopting two distinct values for this parameter: $f_{Ts} = 0.2594$, corresponding to the commonly used value $\sigma_{\pi N} = 55$ MeV, and $f_{Ts} = 0.014$, a value obtained recently from first principles through lattice QCD methods \cite{Dinter:2012tt}. It should be noted that, as clearly stated in Ref.\ \cite{Dinter:2012tt}, this study does not treat systematic uncertainties. It nonetheless provides us with an indicative value for $f_{Ts}$, englobing numerous recent results that point to much lower values than previously estimated.

\newcommand{\Rgaga}{R_{\gamma\gamma}}

\section{Numerical analysis \label{Sec:Analysis}}

\subsection{Setup and strategy}

To analyze which regions of the IDM parameter space are compatible with theoretical and experimental constraints, we perform a numerical analysis scanning randomly over the model parameters. Since the first step of our analysis will involve RGE running of the parameters starting from the input scale (which we take to be the pole mass $M_Z$ of the $Z$-boson), we define the input in the \MSbar\ scheme in the basis of the Lagrangian parameters $\left\{ \lam_2,\lam_3,\lam_4,\lam_5,\mu_2^2 \right\}$, with the exception of the Higgs boson pole mass $M_{\hn}$. For the IDM-specific parameters, we consider the following input ranges:
\begin{align} 
 	0 & < \lam_2(M_Z) < 4\pi, \notag \\ 
  	-4\pi &<  \lam_{3,4,5}(M_Z) < 4\pi, \label{eq:ranges} \\ 
  	-v^2 &< \mu_2^2(M_Z) < \left( 1\,\rm{TeV} \right)^2. \notag
\end{align}
We select the distribution of $\mu_2^2$ in order to obtain a uniform distribution of the CP-even scalar mass $\lm_{\Hn}$. Moreover, in order to improve the sampling efficiency, we generate more points in the regions of parameter space where points are expected to remain valid up to high scales. The  chosen ranges of Eq.~\eqref{eq:ranges} are found to yield both a good parameter space coverage and a sufficiently efficient scan with an acceptable number of viable points, both from the point of view of theoretical constraints at the input scale and with respect to the experimental constraints that we shall impose in the following. The reader should note that the resulting density of points does not have a statistical significance. The existence of a given point is rather to be understood as a parameter set fulfilling certain criteria.

For the first part of the analysis (for which the results are shown in Fig.~\ref{Fig:SMvsIDM}), we do not make any assumption on the value of the Higgs boson mass and let it vary freely within the bounds $0 < M_{\hn} < 500\,\GeV$. We remind the reader that here, and in the following, capital letters always refer to one-loop pole masses. A sample of $10^6$ parameter space points is generated
within the aforementioned ranges, with the only requirement that they fulfill vacuum stability, perturbativity, and unitarity at the input scale $M_Z$.

For the second scan, which is used throughout most of the following analysis, we make the assumption that the LHC experiments have indeed observed a (SM-like) Higgs boson with a mass around $M_{\hn}\simeq 126$~GeV. We therefore fix the mass of the SM-like Higgs boson in the IDM to lie in the range \begin{equation}
\label{HiggsMassRange}
	M_{\hn} = 125.7 \pm 0.6\,\GeV,
\end{equation}
which is obtained by a Gaussian combination of the two experimental measurements \cite{ATLASDiscovery, CMSDiscovery}. This also fixes the value of $\lambda_1$ at the input scale. We then sample $10^7$ points in the IDM parameter space, $\left\{\lam_2,\lam_3,\lam_4,\lam_5,\mu_2^2 \right\}$, again demanding that the stability, perturbativity, and unitarity constraints are satisfied at the scale $M_Z$. 

A further important SM input is the top-quark mass, which in both scans is sampled from the range \cite{Lancaster:2011wr}
\begin{equation}
\label{TopMassRange}
	M_{t} = 173.2 \pm 0.9\,\GeV.
\end{equation}
This mass is also interpreted as the pole mass, related to the running mass at one loop by
\begin{equation}
	\lm_t(\lm_t) = M_t \left[ 1 - \frac{8}{3\pi} \alpha_s(M_t) \right]
\end{equation}

To solve the renormalization group equations numerically, we use a modified version of the {\tt 2HDMC} code (version 1.2) \cite{Eriksson:2009ws, *Eriksson:2010zzb}. At each scale, the conditions for perturbativity, vacuum stability, and unitarity are evaluated. As soon as one of the three conditions fails, we record the corresponding scale, as well as the failing condition. The maximal scale we consider is $Q=10^{16}$~GeV (the ``GUT'' scale), at which the evolution is terminated and surviving points are also recorded. {\tt 2HDMC} is also used to evaluate the oblique parameters (at the input scale), and we apply the bounds from direct searches at colliders as described in Sect.~\ref{Sec:collider}. For each parameter space point, we evaluate the scalar pole masses from the running parameters as described in Sec.\ \ref{Sec:DiagrammaticCalculation}. This set of parameters is passed to \MO, which is used to compute the relic density and the WIMP-nucleon spin-independent scattering cross section. As a final step, we compute the Higgs decay rates (again using {\tt 2HDMC}) to check whether a modification of the important $\hn \rightarrow \gamma \gamma$ decay mode could be reproduced in any of the viable parameter points, and to apply constraints from invisible Higgs decays into LOP pairs.

\subsection{Extrapolation scale}

\begin{figure}
	\begin{center}
		\includegraphics[width=0.49\columnwidth]{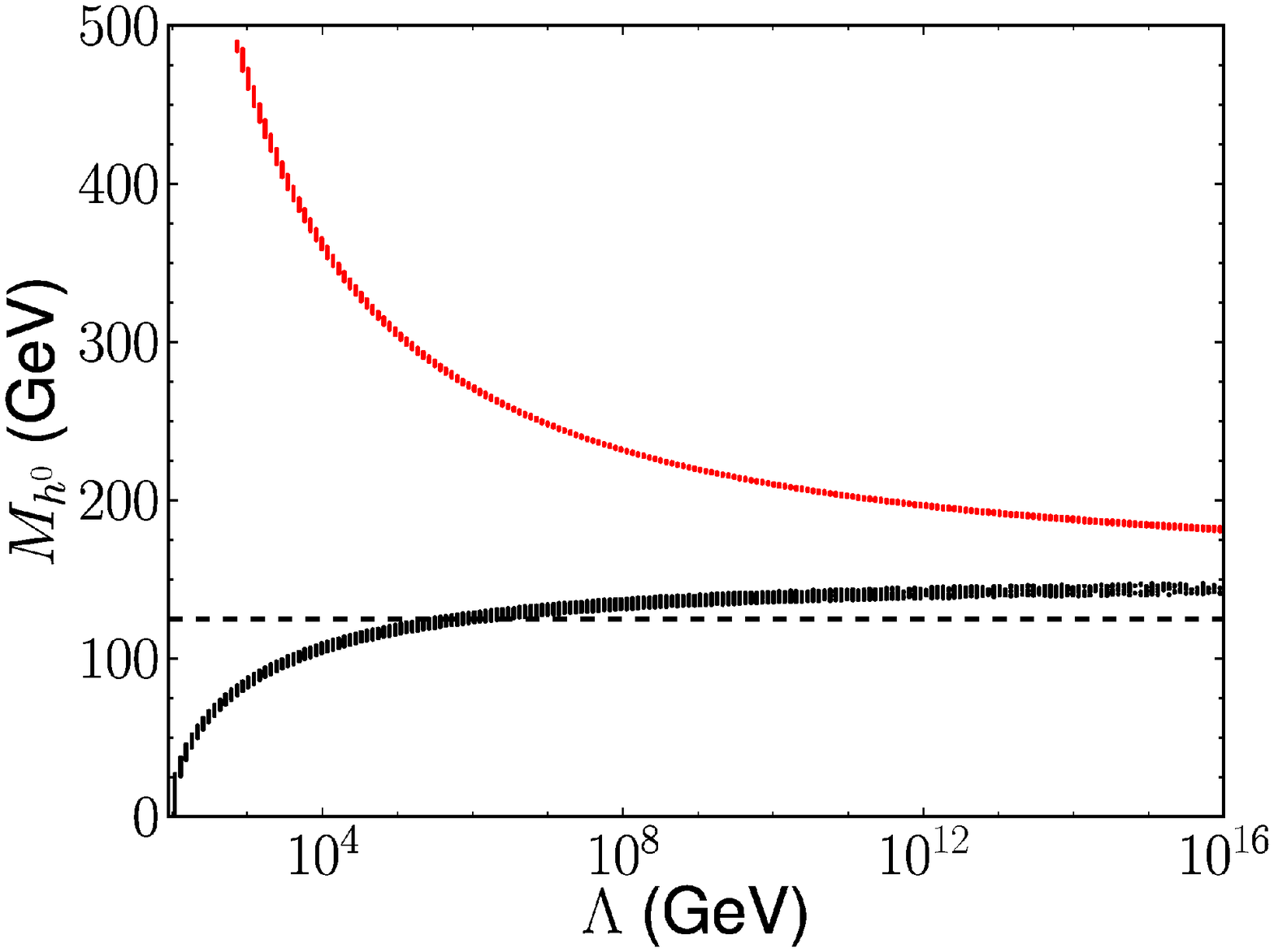}
		\includegraphics[width=0.49\columnwidth]{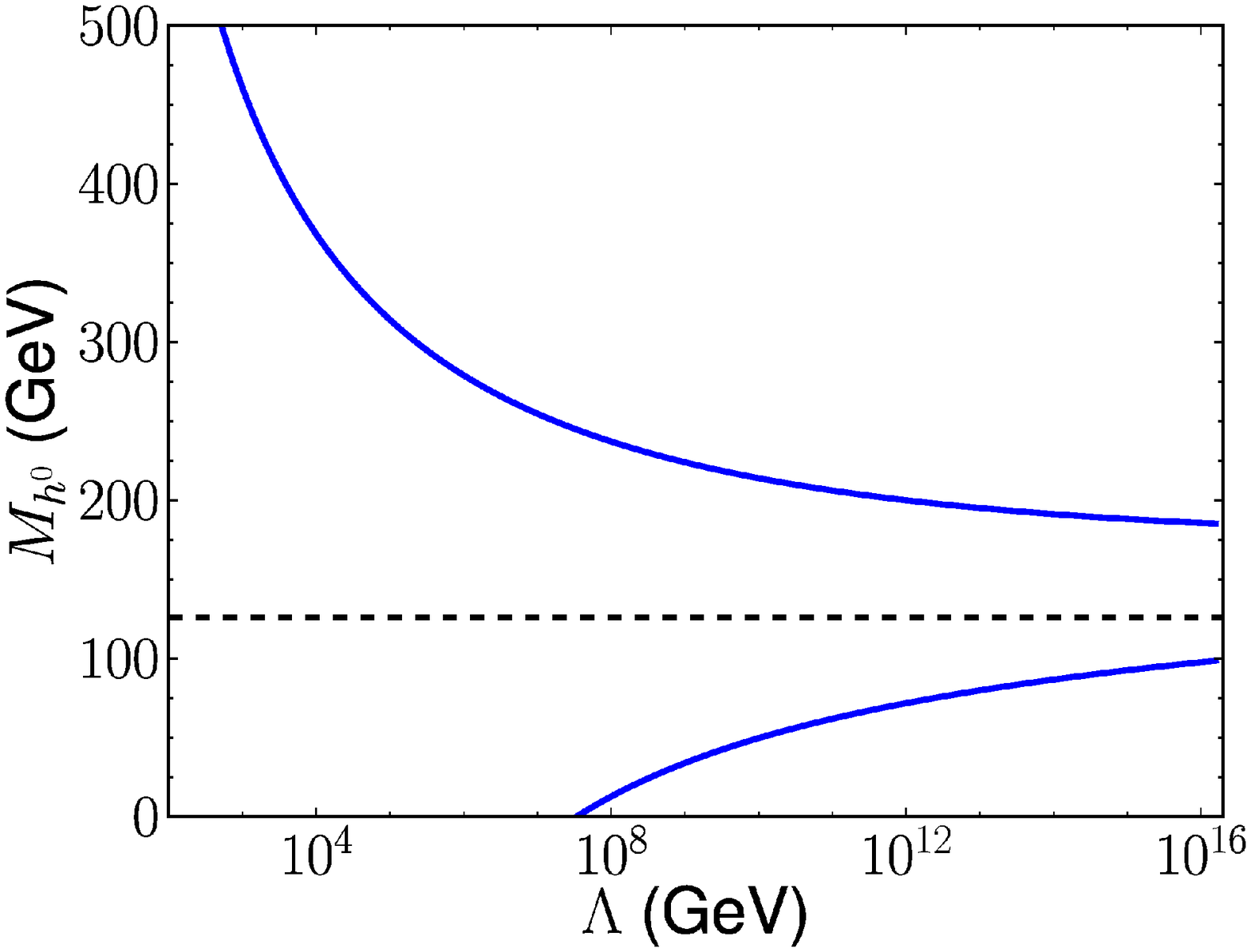}
	\end{center}
	\vspace{-1em}
	\caption{Left: Results from a one-loop RGE analysis of the maximum extrapolation scale $\Lambda$ which satisfies perturbativity (red), stability (black), and unitarity (not relevant) as a function of the Higgs mass $M_{\hn}$ in the Standard Model. The value corresponding to the LHC signal is indicated by a dashed line. Right: The corresponding results in the Inert Doublet Model for the parameter variations according to Eq.\ \eqref{eq:ranges}.}
	\label{Fig:SMvsIDM}
\end{figure}

Our first goal is to examine the maximal scale $\Lambda$ up to which the IDM can be extrapolated while remaining a consistent and calculable quantum field theory, in the sense of preserving perturbativity of all couplings, unitarity of the scattering matrix, and (absolute) stability of the electroweak vacuum as discussed above. 

We first employ the scan with variable Higgs mass. For comparison, we also perform the same excercise for the SM, where the only free parameter is taken to be the mass of the Higgs boson $\hn$. Both analyses are performed on the same footing, using one-loop renormalization group equations and varying the top quark mass within the limits of Eq.\ \eqref{TopMassRange}.\footnote{Far more sophisticated analyses exist for the SM, where state-of-the-art calculations rely on three-loop RGEs for the couplings \cite{EliasMiro:2011aa}. This analysis shows that the SM Higgs mass range yielding an absolutely stable vacuum is enlarged by a few GeV compared to the lowest order prediction.}
The results are presented in Fig.~\ref{Fig:SMvsIDM}, where we show the maximal extrapolation scale of the SM (left) and the IDM (right) as a function of $M_{\hn}$. The value corresponding to the LHC observation, as given by Eq.~\eqref{HiggsMassRange}, is indicated by the dashed line. 

In the left-hand side plot of Fig.~\ref{Fig:SMvsIDM}, we can see the two well-known SM bounds coming from perturbativity (red), and vacuum stability (black), constraining the Higgs mass from above and below, respectively. The width of these two lines corresponds to the uncertainty from variation of the top quark mass. It should be noted that the Higgs mass favored by the LHC discovery does not satisfy absolute stability in our simple one-loop analysis. A more sophisticated analysis points towards a (long-lived) metastable vacuum for the SM \cite{EliasMiro:2011aa}.
 
The introduction of the additional inert doublet significantly modifies the extrapolation constraints as a function of the Higgs mass, as can be seen on the right-hand side of Fig.~\ref{Fig:SMvsIDM}. While the modification of the upper bound from perturbativity is less pronounced, the lower one from stability is greatly relaxed, with the electroweak vacuum being able to satisfy absolute stability up to high scales for a larger choice of the higgs mass than in the SM case. This effect is due to the contributions of the inert scalars to the potential, that can counteract the tendency of the top quark Yukawa coupling to destabilize the vacuum by driving $\lambda_1$ negative, see Refs.~\cite{Nie:1998yn,Kanemura:1999xf}. Interestingly, this leads to a vacuum stability constraint that only manifests itself fairly early in the RGE evolution. In other words, parameter space points which have a stable vacuum at the input scale are likely to remain stable up to high scales. It is therefore generally the case that one of the other two requirements (perturbativity or unitarity) is the more constraining for $M_{\hn}$ (as can be seen From Fig.~\ref{Fig:SMvsIDM}).
We find that the range for which the IDM Higgs mass can be extrapolated up to the GUT scale is larger than that in the SM, and in particular that lower values for $M_{\hn}$ are allowed. The measured value $M_{\hn} \sim 126\GeV$, which at our level of approximation does not allow for the SM to be valid up to the GUT scale, see Fig.~\ref{Fig:SMvsIDM} (left), is therefore allowed in the IDM. We expect this conclusion to hold also in a more sophisticated analysis since, as already mentioned, the SM bound on $M_{\hn}$ from stability is \emph{relaxed} by the inclusion of higher-order corrections \cite{EliasMiro:2011aa}. Futhermore, allowing for metastability would render an even larger IDM parameter space allowed, as is known to occur for the SM. A more detailed analysis of this constraint is beyond the scope of the present work.
\\

\begin{figure}
	\begin{center}
		\includegraphics[width=0.95\columnwidth]{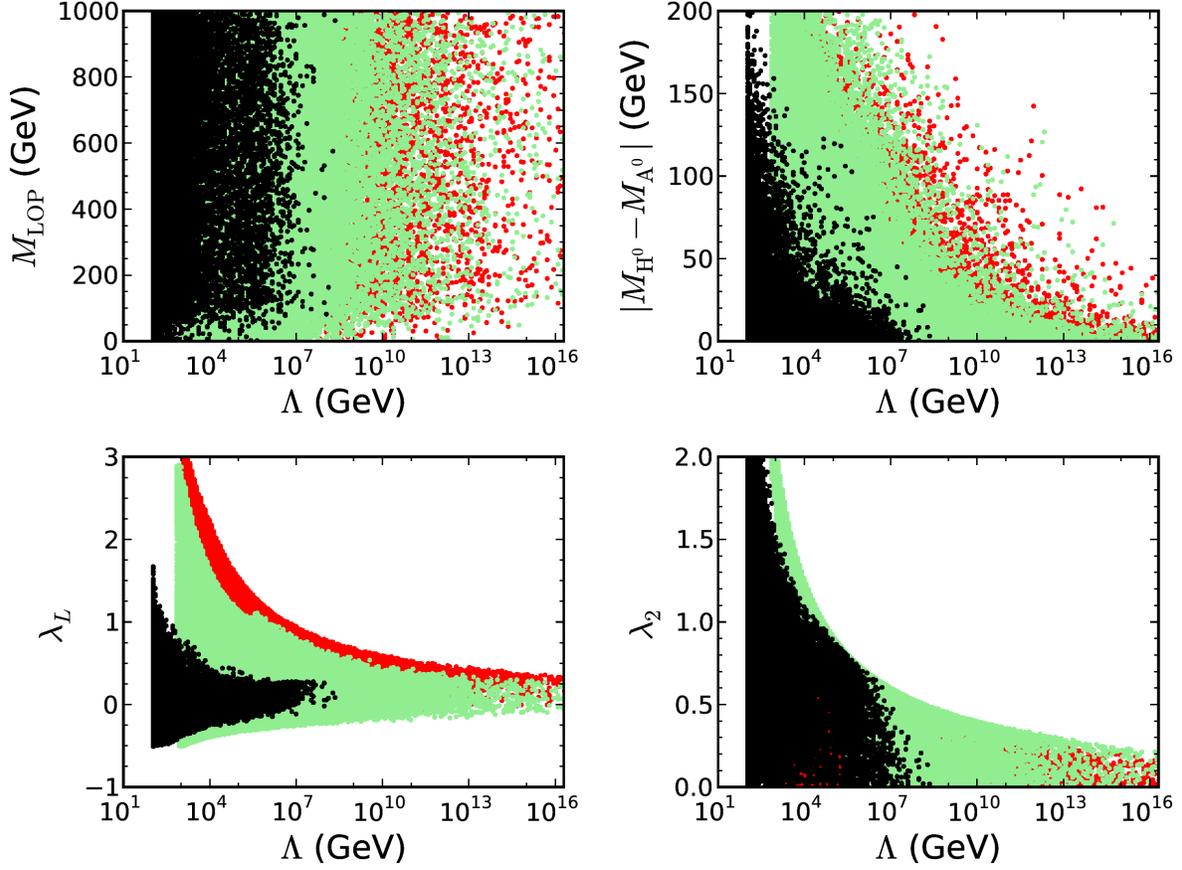}
		\vspace{-1.5em}
	\end{center}
	\caption{Maximum scale $\Lambda$ up to which the IDM satisfies perturbativity, vacuum stability, and unitarity as a function of different model parameters/predictions: the LOP mass (top left), the mass difference $|M_{\Hn}-M_{\An}|$ (top right), the $\Hn-\Hn-\hn$ coupling $\lambda_L$ (bottom left) and the $\lambda_2$ parameter (bottom right). The criterion which failed at the lowest scale is indicated by the colour code: perturbativity (red points), vacuum stability (black), and unitarity (green).}
	\label{Fig:SPUconstraints}
\end{figure}
In the following we constrain the Higgs boson mass to the range of Eq.\ \eqref{HiggsMassRange} suggested by the LHC discovery. We also impose the oblique parameter and collider constraints presented in Secs.\ \ref{Sec:oblique} and \ref{Sec:collider}. As before, we evolve the RGEs until one of the three theoretical requirements fails and record the corresponding scale. The results are presented in Fig.~\ref{Fig:SPUconstraints}, where the scale $\Lambda$, at which the first of the constraints fails, is presented against four different parameters/predictions of the model, namely $\Mlop$, $|M_{\Hn}-M_{\An}|$, $\lambda_L$, and $\lambda_2$. The colour coding is the same as in Fig.~\ref{Fig:SMvsIDM}: red points fail due to the perturbativity constraint, black points fail due to vacuum stability, and the green points are those for which unitarity is violated at the scale $\Lambda$. We observe that, as also noted previously, the vacuum stability constraint is relevant mostly at rather low scales $\Lambda \lesssim 10^7$ GeV. Unitarity violation can occur as late as the GUT scale. 

From Fig.\ \ref{Fig:SPUconstraints} we also learn that extrapolating the model towards the GUT scale does not favour a particular range for the mass of the dark matter candidate (the LOP). However, the coupling parameters are strongly constrained when reaching scales close to the GUT scale. This is due to the requirement of perturbativity, which is much easier to fulfill if the numerical values of the couplings are small. As a further consequence, the $\Hn-\Hn-\hn$ coupling $\lambda_L$ and the four-scalar coupling $\lambda_2$ have to be chosen in a rather narrow interval for the model to be valid up to the GUT scale.
The mass difference between $\Hn$ and $\An$ is also limited if the GUT scale validity requirement is applied. This is due to the fact that $\lm_{\Hn}$ and $\lm_{\An}$ stem from a common mass scale, $\mu_2$, and their difference is proportional to $\lambda_5 v^2$ (see Eqs.\ \eqref{Eq:mH0tree}). Since the quartic couplings are bound due to the perturbativity requirement, the mass difference for valid points naturally decreases when approaching the GUT scale.
 \looseness=-1

\begin{figure}
\vspace{-2em}
	\begin{center}
		\includegraphics[width=0.9\columnwidth]{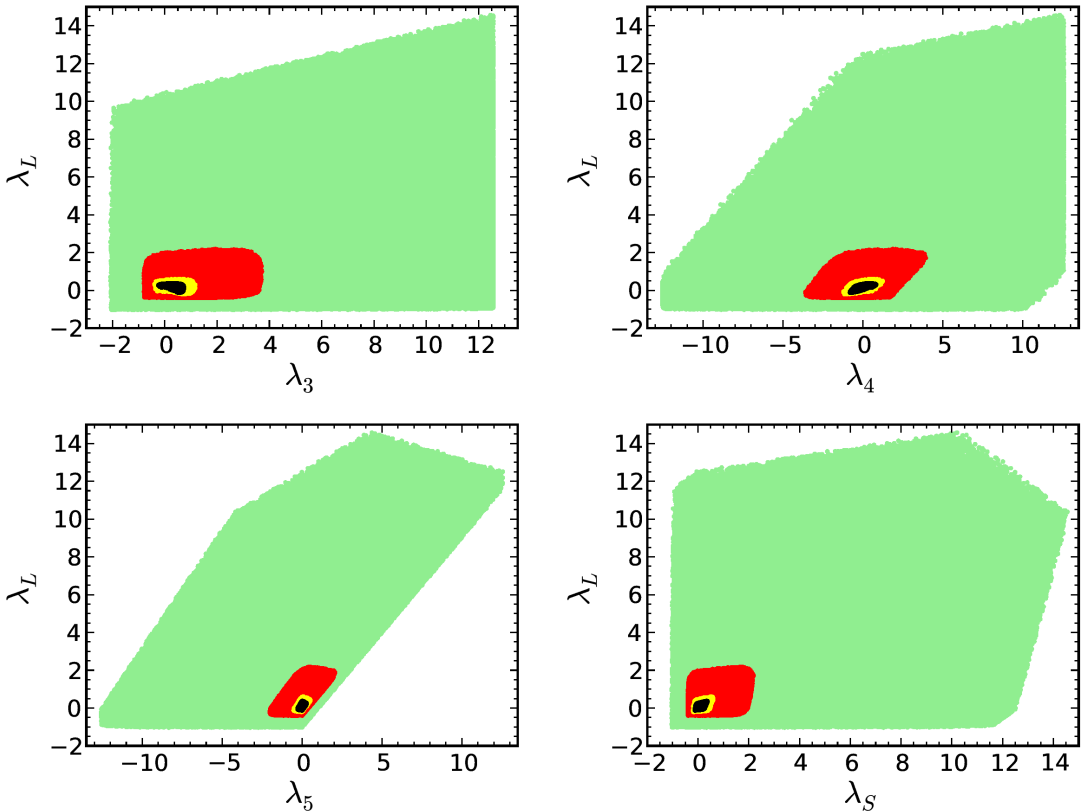}
	\end{center}
	\vspace{-2em}
	\caption{Projections of the parameter space of the Inert Doublet Model on the planes of $\lambda_L$ against $\lambda_3$, $\lambda_4$, $\lambda_5$, and $\lambda_S$. The green regions correspond to all valid points in the scan at the input scale $\Lambda=M_Z$, while the red, yellow, and black regions show the points which remain valid up to $\Lambda=10^4$ GeV, $\Lambda=10^{10}$ GeV, and the GUT scale $\Lambda=10^{16}$ GeV, respectively.}
	\label{Fig:lamL}
\end{figure}
\begin{table}
\centering
\begin{tabular}{|c|c|cccc|}
\hline
~Parameter~ & ~Scan range~ & \multicolumn{4}{|c|}{Valid range} \\
              & ~$\Lambda=M_Z$~  & ~$\Lambda=M_Z$~ & ~$\Lambda=10^4$~GeV~ & ~$\Lambda=10^{10}$~GeV~ & ~$\Lambda=10^{16}$~GeV~ \\  
\hline
$\lambda_2$& $(0, 4\pi )$& $(0, 8.4)$ & $(0, 1.30)$ & $(0, 0.40)$ & $(0, 0.15)$\\
$\lambda_3$& $(-4\pi, 4\pi)$& $(-2.2, 4\pi)$ & $(-0.80, 3.60)$ & $(-0.40, 1.15)$ & $(-0.25, 0.75)$\\
$\lambda_4$& $(-4\pi, 4\pi)$& $(-4\pi, 4\pi)$ & $(-3.35, 3.70)$ & $(-1.20, 1.25)$ & $(-0.80, 0.90)$ \\
$\lambda_5$& $(-4\pi, 4\pi)$& $(-4\pi, 4\pi)$ & $(-1.95, 1.95)$ & $(-0.55, 0.55)$ & $(-0.30, 0.30)$ \\
$\lambda_L$& -- & $(-1.2, 14.6)$ & $(-0.40, 2.20)$ & $(-0.20, 0.60)$ & $(-0.15, 0.40)$ \\
$\lambda_S$& -- & $(-1.2, 14.6)$ & $(-0.40, 2.20)$ & $(-0.20, 0.60)$ & $(-0.15, 0.40)$ \\
\hline
\end{tabular}
\caption{Input ranges for the IDM quartic couplings, allowed ranges when imposing constraints at the input scale, and after extrapolation to the scales of $10^4$ GeV, $10^{10}$ GeV, and the GUT scale $10^{16}$ GeV.}
\label{Tab:lambdas}
\end{table}
The same behaviour is observed for the remaining coupling parameters, as shown in Fig.~\ref{Fig:lamL} where the IDM parameter space is projected on the planes $(\lambda_3, \lambda_L)$, $(\lambda_4, \lambda_L)$, $(\lambda_5, \lambda_L)$, and $(\lambda_S, \lambda_L)$. The (light) green regions in this figure correspond to parameter space points that pass the aforementioned constraints at the input scale $M_Z$ (constraints from DM observables are not imposed here; they will be discussed in the next section). Thus, they constitute the maximal phenomenologically
viable regions of the IDM within the framework of our one-loop analysis.

At the same time, we highlight in different colours the points corresponding to parameter values for which the IDM can be extrapolated up to $\Lambda=10^4$ GeV (red), $\Lambda=10^{10}$ GeV (yellow), and the GUT scale $\Lambda=10^{16}$ GeV (black) (assuming that there is no new physics between the input and the extrapolation scale). As can be seen from this figure, the constraints imposed at the input scale are not too prohibitive, and they are satisfied for essentially the full parameter ranges considered in our scan. It can also be seen that the IDM can be extrapolated up to the GUT scale ($\Lambda=10^{16}$ GeV) in sizeable regions of the parameter space, with the ranges of $\lambda_{L}$, $\lambda_{S}$, and $\lambda_5$ for which this can happen being somewhat more limited, whereas $\lambda_{3}$ and $\lambda_4$ can reach higher absolute values. Considering new physics entering at a lower scale, e.g., at $\Lambda=10^4$ GeV or $\Lambda=10^{10}$ GeV, the allowed parameter ranges increase, cf.~Fig.~\ref{Fig:SPUconstraints}.
Note, however, that the valid ranges for the coupling parameters at the phenomenologically relevant scale $\Lambda=10^4~{\rm GeV}=10~{\rm TeV}$ is already rather reduced with respect to the input scale $\Lambda=M_Z$.
In Tab.~\ref{Tab:lambdas} we summarize the allowed ranges for the quartic couplings, both at the input scale and after the evolution to the mentioned scales.

\subsection{Dark Matter}

\begin{figure}[b!]
	\begin{center}
		\includegraphics[width=\columnwidth]{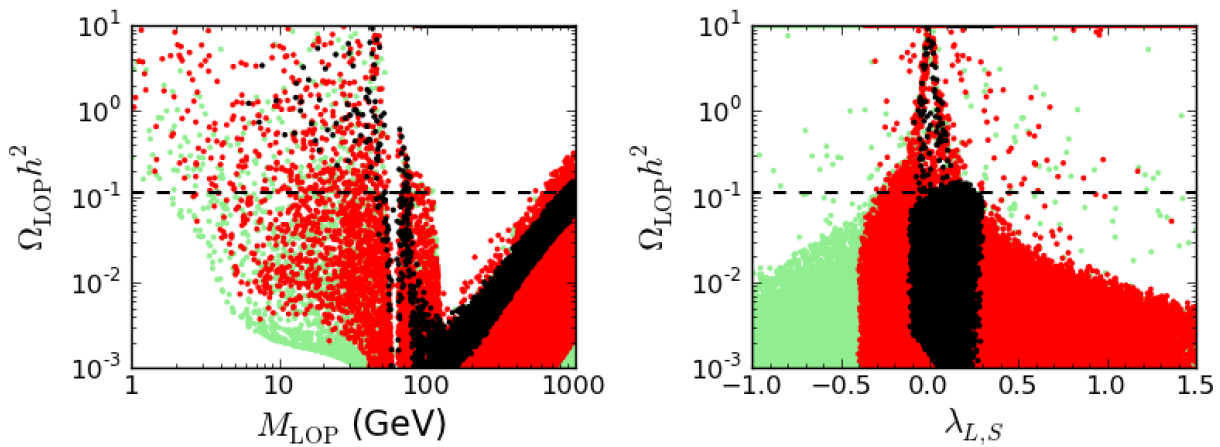}
	\end{center}
	\vspace{-2em}
	\caption{The dark matter relic density versus the LOP mass $M_{\rm LOP}$ (left) and the coupling of a LOP pair to the Higgs boson $\lambda_{L,S}$ (right). The green points correspond to all valid points in the scan at the input scale $\Lambda=M_Z$, while the red and black region show the points which remain valid up to the scale $\Lambda=10^4$ GeV and the GUT scale $\Lambda=10^{16}$ GeV, respectively. The black dashed line indicates the WMAP central value of Eq.~\eqref{Eq:RelicDensity}.}
	\label{Fig:RelicDensity1}
\end{figure}

With the previous results at hand, we now turn to the IDM dark matter phenomenology, assuming as previously that the Higgs and top quark mass are fixed within the ranges given by Eqs.~\eqref{HiggsMassRange} and \eqref{TopMassRange}, respectively. In Fig.~\ref{Fig:RelicDensity1} we show the predicted relic density as a function of the DM candidate mass, $M_{\rm LOP}$ (left), and the triple coupling of a DM pair to the (SM-like) Higgs boson ($\lambda_L$ when the LOP is $\Hn$, $\lambda_S$ when it is $\An$) (right). The $3\,\sigma$ limit from the $7$-year WMAP data is represented by the red-dashed region.\footnote{An updated result $\Omega_{\rm CDM}h^2 ~=~0.1196  \pm 0.0031$ has recently been obtained by the Planck collaboration~\cite{Ade:2013zuv}. As is clear from Fig.~\ref{Fig:RelicDensity1}, using this value would not lead to a qualitative difference in our results. In particular the resulting upper limit on the DM density remains numerically very similar, see Eq.~\eqref{Eq:RelicDensityUpper}.} 

From the left-hand side plot in Fig.~\ref{Fig:RelicDensity1}, we can see that the correct relic density can be achieved in the mass regimes that we described in Sec.~\ref{Sec:IDMoneloop}. The viable parts of the low- and intermediate- mass regimes extend from $3$ GeV up to roughly $120$ GeV. From this value and up to approximately $500$ GeV, the predicted relic density is too low and it can reach the WMAP levels again above $500$ GeV. In the right-hand side plot, we see that the values that $\lambda_L$ ($\lambda_S$) can take while yielding the correct relic density lie in the range $-0.4 \lesssim \lambda_{L,S} \lesssim 0.4$, with positive values preferred by the points viable up to the GUT scale.

\begin{figure}
	\begin{center}
		\includegraphics[width=\columnwidth]{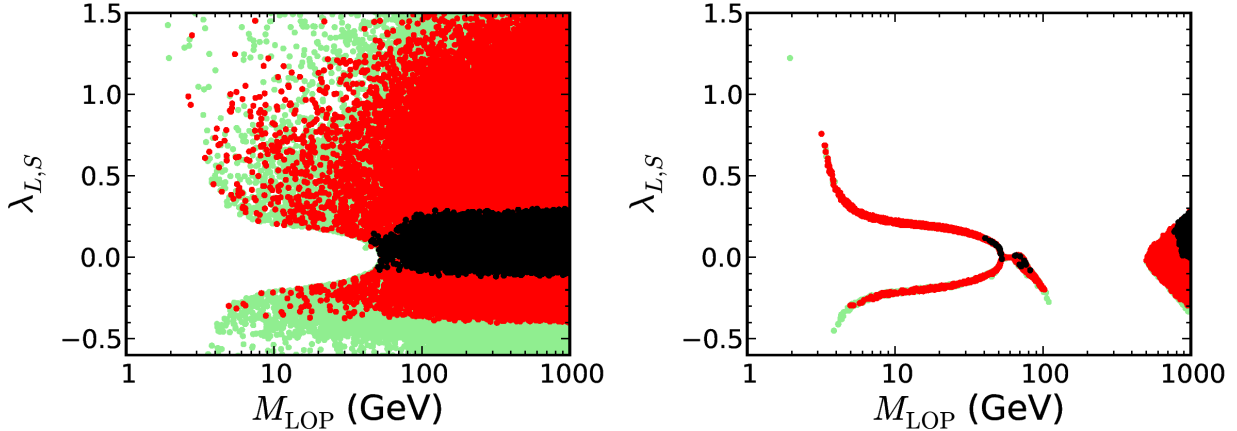}
	\end{center}
	\vspace{-2em}
	\caption{The viable IDM parameter space projected on the $(M_{\rm LOP},\lambda_{L,S})$ plane imposing only the upper limit (left) and the upper and lower limits (right) of the WMAP range of Eq.~\eqref{Eq:RelicDensity}. The colour coding is the same as in Fig.~\ref{Fig:RelicDensity1}.}
	\label{Fig:RelicDensity2}
	\vspace{-1em}
\end{figure}

In order to better illustrate the impact of the WMAP results, in Fig.~\ref{Fig:RelicDensity2} we project the IDM parameter space onto the $(M_{\rm LOP}, \lambda_L)$ plane, demanding that the relic density satisfies only the upper $7$-year WMAP limit (left), or both the upper and the lower bound (right).
In particular the right plot shows how restrictive the full DM constraint is, and it allows to discuss in more detail the various mechanisms responsible for producing the correct values for $\Omega_{\mathrm{LOP}}h^2$ in the different mass regimes.
 
At very low masses, below 4 GeV, LOPs annihilate dominantly into $\tau$ pairs through $s$-channel Higgs exchange. The most energetic part of the LOP's Boltzmann velocity distribution can also give some annihilation into bottom quarks. The correct relic density is achieved by adjusting $\lambda_{L,S}$ which, given the relative smallness of the $\tau$ Yukawa coupling, has to be fixed at rather large values. These values decrease as phase space opens up, but also as a larger part of the WIMP Boltzmann distribution passes the $b\bar{b}$ production threshold. Above $\Mlop \sim 4$ GeV, the required $\lambda_{L,S}$ values drop by a factor $3$--$4$, compensating for the larger bottom Yukawa coupling. Continuing up to roughly $50$ GeV, the bulk of the points lie on a thin strip where $\lambda_{L,S}$ decreases mildly, corresponding to dominant DM annihilation into $b\bar{b}$ pairs through Higgs boson exchange. We remind the reader that due to the LEP direct search constraints, coannihilation is absent in this mass range. For $\Mlop \sim 50$ GeV, we observe that the WMAP-allowed values for $\lambda_{L,S}$ decrease further, approaching $\lambda_{L,S}=0$ at precisely $M_{\rm LOP} = M_{\hn}/2$. Two effects are at play in this mass region. First, the LOP mass approaches the point where the Higgs propagator becomes resonant. For a constant $\lambda_L$ value, the LOP self-annihilation cross section then increases dramatically. As a result, in order to get the correct relic density, $\lambda_{L,S}$ must attain small values. The second important effect in this mass region comes from the contributions of virtual gauge boson final states, and especially from $W^{(*)}W^*$. Even for vanishing triple coupling of a DM pair to $\hn$, there is an additional four-vertex involving two DM particles and two $W$ bosons. This results from the covariant derivative terms in the Lagrangian, acting on the $\Phi$ doublet (minimal coupling). The strength of this vertex is thus characterized by a gauge coupling, and does not depend on any free parameter of the model (see also Ref.~\cite{Honorez:2010re}). This constitutes a distinct positive contribution to the DM annihilation cross section that pushes the required value for $|\lambda_{L,S}|$ further down. 

For the IDM high-mass regime, a detailed explanation for the behaviour of the the relic density constraint has been presented in Ref.~\cite{Hambye:2009pw}. Above roughly 120 GeV (the precise value depends on $M_{\hn}$ \cite{LopezHonorez:2010tb}), the cross section for self-annihilation, in particular into gauge boson final states, becomes prohibitively large for the lower WMAP bound to be satisfied. We therefore see that, whereas this region is allowed in the sense of not overshooting the WMAP limit, Figs.~\ref{Fig:RelicDensity1}, \ref{Fig:RelicDensity2} (left), it fails once we demand the IDM to fully account for the measured DM density of the universe, Fig.~\ref{Fig:RelicDensity2} (right). However, for $\Mlop>500$ GeV, an interesting effect comes at play. The dominant contribution to the cross section of the type $\Hn \Hn \rightarrow VV$ ($\An \An \rightarrow VV$) here comes from the longitudinal gauge boson components. This cross section is mediated either by a direct quartic coupling, or by a $t$/$u$-channel exchange of $\Hn$, $\An$, or $\Hp$, which scales as $M_{\Hn}^2/M_Z^2$ (in the case of the $Z$ final state, and similarly for the $W$). The annihilation cross section therefore becomes very large as $\Mlop$ increases (with a similar dependence also on the NLOP mass). However, when the $\Hn$ and $\An/\Hp$ are nearly mass-degenerate, there is a cancellation taking place between the $t$/$u$ channel contributions and the four-vertex diagram. This cancellation is exact for an exact degeneracy. For example, with $\lambda_L = 0$, $M_{\Hn} = 700$ GeV, and $M_{\An} = 701$, the relic density would be too high, and to satisfy the WMAP bound a non-zero value for $\lambda_L$ is required. In this manner, the WIMP depletion rate can be balanced by varying the LOP-NLOP mass splitting and the $\lambda_L$ parameter to obtain the correct mixture of transverse and longitudinal gauge bosons in the final state. These solutions are always found for small LOP-NLOP mass splittings, and require some tuning of the value of $\lambda_{L,S}$. In practice we find that the maximal allowed mass splitting for the points in our scan is of the order $10$ GeV.

\begin{figure}[b!]
\vspace{-1em}
	\begin{center}
		\includegraphics[width=\columnwidth]{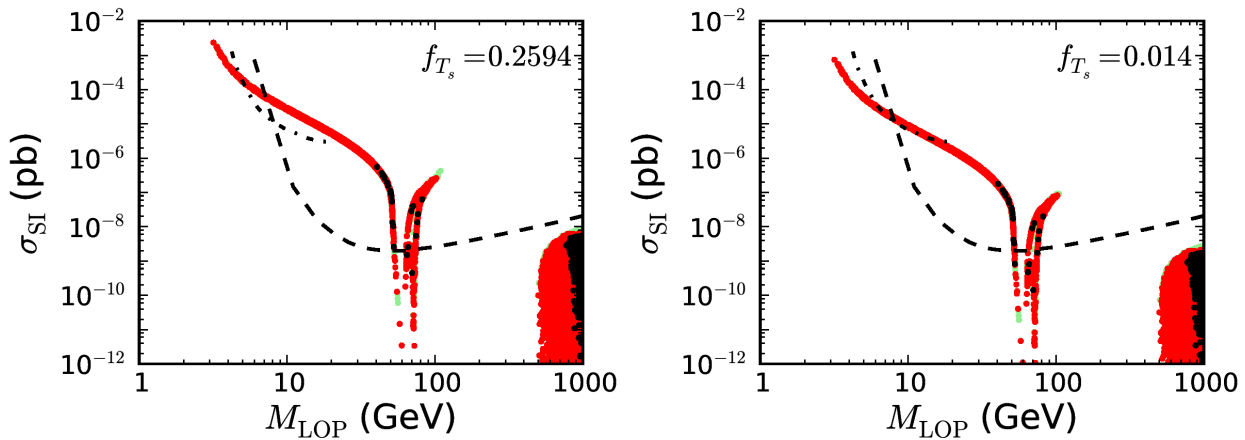}
	\end{center}
	\vspace{-2em}
	\caption{Projection of the viable IDM parameter space on the $(M_{\rm LOP},\sigma_{\rm SI})$ plane against the
        latest limits from XENON10 (dash-dotted line) and XENON100 (dashed line) for $f_{Ts} = 0.2594$ (left) and $f_{Ts} = 0.014$ (right). The colour coding for the shown points is the same as in Fig.\ \ref{Fig:RelicDensity1}.}
	\label{Fig:DirectDetection}
\end{figure}

Further constraints from dark matter come, as we have already mentioned, from direct detection experiments, and most notably the latest XENON100 limits on the WIMP-nucleon spin-independent scattering cross section, $\sigma_{\rm SI}$. In Fig.\ \ref{Fig:DirectDetection} we show the viable IDM parameter space in the usual $(M_{\rm LOP},\sigma_{\rm SI})$ space and confront the model to the latest exclusion bounds from XENON100 \cite{Aprile:2012nq} (dashed line) and XENON10 \cite{Angle:2011th} (dash-dotted). As discussed in Section~\ref{sect:direct}, we adopt two distinct values for the strange quark nucleon form factor, and the results are shown in the left- and right panels of Fig.\ \ref{Fig:DirectDetection}, respectively. In this figure all points respect both the upper and the lower WMAP bounds. The behaviour of the LOP-nucleon scattering cross section follows quite closely the corresponding behaviour of the coupling $\lambda_{L,S}$ in Fig.~\ref{Fig:RelicDensity2}, since the only way of coupling the LOP to quarks at tree level is through $t$-channel Higgs exchange (the relevant coupling being simply $\lambda_{L,S}$). It can be seen clearly that  the low- and intermediate- mass regimes are almost fully excluded. The only surviving points are those for which the correct relic density is achieved through a combination of small values for the $\lambda_{L,S}$ coupling and quasi-resonant annihilation to an $s$-channel Higgs boson and virtual gauge boson final states. The very low mass regime ($\Mlop < 10\GeV$) is also excluded by the XENON10 bounds on low-mass WIMPs \cite{Angle:2011th}. The high mass regime, on the other hand, remains unaffected by current direct detection bounds.

In Fig.~\ref{Fig:DirectDetection} we can also clearly observe a consequence of fixing the Higgs mass to a constant value. The resonance regime, which for a variable Higgs mass consists of an entire region, acquires now the well-known two-branch ``funnel'' structure that is present, e.g., in models such as the MSSM, where the Higgs mass is fixed to a very limited range. The funnel extends quasi-symmetrically in two narrow regions around $M_{\hn}/2$. The LOP cannot approach too close to the resonance since the relic density would then be too low. 
With the Higgs mass fixed, the freedom of moving within the funnel region essentially disappears, which demonstrates that a greater accuracy is demanded in order to assess whether a parameter space point is excluded or not. A similar comment applies to the high mass regime and the precise tuning of the LOP-NLOP mass splitting that we discussed above.

Moreover, we observe the anticipated, numerically significant, impact of $f_{Ts}$ on the LOP-nucleus scattering cross section. As we pointed out, the driving mechanism for WIMP-nucleon scattering in the IDM is $t$-channel Higgs exchange. The Higgs couples preferentially to strange quarks, hence varying the strange quark form factor can have an impact of up to a factor $7$ in the scattering cross section. While in this particular case most of the parameter space remains excluded, this is a transparent illustration of the importance of knowing $f_{Ts}$ with better accuracy to make precise statements. Using the more conservative value, we find that the mass ranges allowed by WMAP and direct exclusion bounds are $50\GeV< \Mlop <  80\GeV$ and $\Mlop>500\GeV$.

As a final remark on direct detection, we should mention that in a recent paper the one-loop electroweak corrections to the LOP-nucleon scattering cross section were computed \cite{Klasen2013EWDD}. According to the findings, these corrections give positive contributions to the scattering cross section of the order of $10^{-11}$--$10^{-10}$ pb, that can actually superseed the tree-level contributions in non-negligible regions of the parameter space (in particular for small $\lambda_L$ values). These contributions do not depend on the IDM couplings, since the relevant vertices involve gauge couplings, and for the most should be quite insensitive to variations of $f_{Ts}$. The most striking impact is that these corrections basically set a lower value for the scattering cross section over the full range of LOP masses, which turns out to be within the reach of the upcoming XENON-1T experiment. This is a nice example of the complementarity among direct dark matter and collider searches, since the high mass regime, most probably out of the LHC reach, will be probed by the next generation of direct detection experiments.

\subsection{LHC Higgs phenomenology}

The phenomenology of the Higgs-like state observed at $M_{\hn} \sim 126\GeV$ remains mostly unchanged in the IDM compared to the SM. Since the inert scalars do not have tree-level couplings to fermions, the SM production modes and dominant decay channels are unaffected, which leaves the rates compatible with the data from different channels measured by ATLAS \cite{ATLASDiscovery} and CMS \cite{CMSDiscovery}. On the other hand, the prediction for one of the experimentally most important decay modes, $\hn \to \gamma\gamma$, can be affected at leading order. In the SM, the $\hn \to \gamma\gamma$ decay is mediated by loop diagrams, where the most important contributions come from $W$ bosons and top quarks, while in the IDM it receives an additional contribution from the charged scalars.\footnote{The leading-order prediction for the decay $\hn \to Z\gamma$ could be similarly modified, but this mode has not yet been observed experimentally.} Following the LHC discovery, this has also been discussed in the context of the 2HDM \cite{Ferreira:2011aa, *Ferreira:2012my, *Arhrib:2012ia, *Altmannshofer:2012ar, *Swiezewska:2012eh}.

\begin{figure}
	\begin{center}
		\includegraphics[width=0.95\columnwidth]{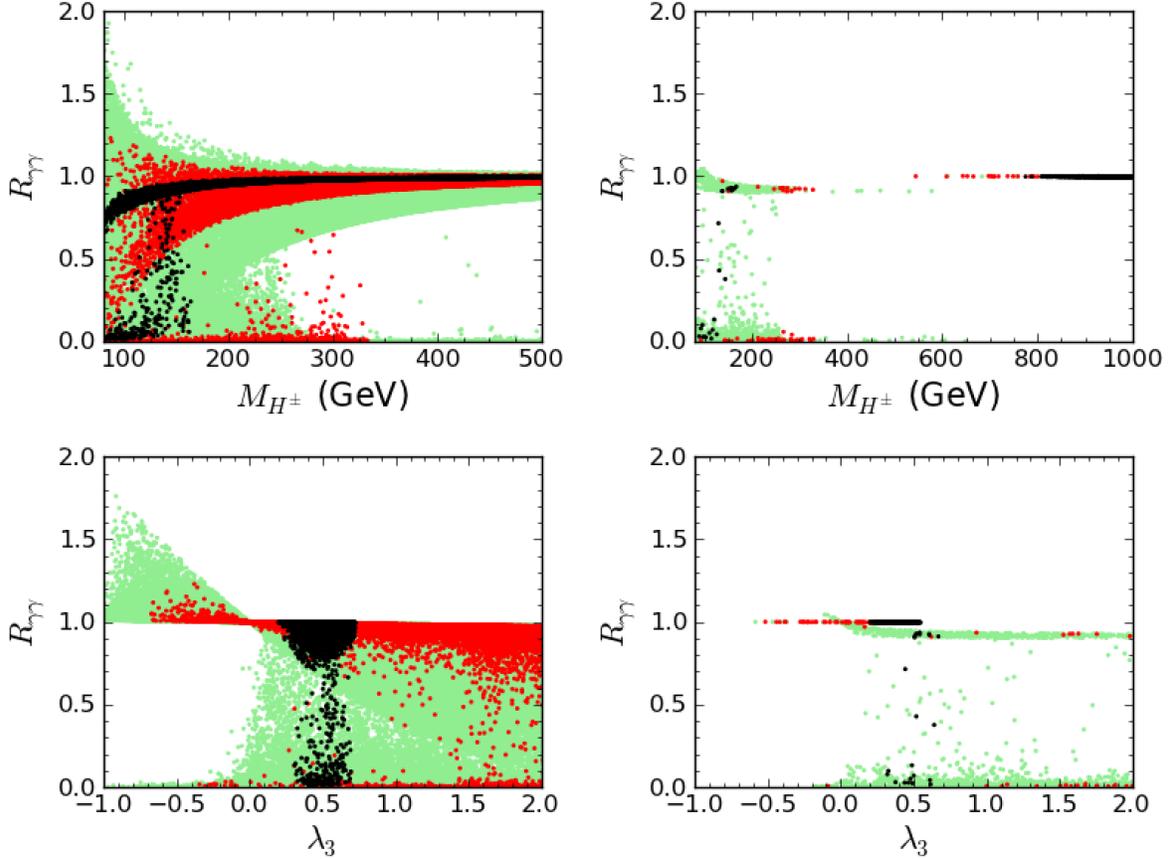}
	\end{center}
	\vspace{-2em}
	\caption{Relative branching ratio $\Rgaga=\mathrm{BR}(\hn \rightarrow \gamma \gamma)_{\mathrm{IDM}}/\mathrm{BR}(\hn \rightarrow \gamma \gamma)_{\mathrm{SM}}$ as a function of the charged scalar mass $M_{\Hp}$ (top) and the coupling $\lambda_3$ (bottom), applying only the upper WMAP limit on the relic density (left), and both the upper and lower limits (right). The colour coding is the same as in Fig.~\ref{Fig:RelicDensity1}.}
	\label{Fig:hgaga}
\end{figure}

We calculate the partial width $\Gamma(\hn\to \gamma\gamma)$ and the corresponding branching ratio for the points in the parameter space that satisfy the constraints, including the WMAP bounds on the relic density. In Fig.~\ref{Fig:hgaga}, we show the ratio
\begin{equation}
	\Rgaga = \frac{\mathrm{BR}(\hn \rightarrow \gamma \gamma)_{\mathrm{IDM}}}{\mathrm{BR}(\hn \rightarrow \gamma \gamma)_{\mathrm{SM}}},
\end{equation}
which in the IDM corresponds to the modification of the inclusive $\hn\to \gamma\gamma$ rate (recall that the $\hn$ production in, e.g., gluon fusion, remains unchanged). The colour coding used in Fig.~\ref{Fig:hgaga} is the same as above, where green indicates all viable points in the scan at the input scale $\Lambda=M_Z$, red corresponds to the points which can be extrapolated to at least $\Lambda=10^4$ GeV, and the black points in addition remain stable and perturbative up to the GUT scale $\Lambda=10^{16}$ GeV.
Considering first the upper left plot, we show $\Rgaga$ as a function of $M_{\Hp}$ with only the upper WMAP limit on the relic density applied. 
For large values of $M_{\Hp}$ the rate decouples to the SM value ($\Rgaga\to 1$), whereas for $M_{\Hp}\lesssim 300\GeV$, we see that the two photon rate can be modified substantially. Here we observe scenarios with an enhancement in $\Rgaga$ above the SM, or with a suppression to lower values. The effect is most pronounced for low $M_{\Hp} \lesssim 100\GeV$, where the possible modification due to scalar contributions exceeds $30\%$. It should be pointed out that these low values for the charged scalar mass are allowed in the IDM without contradicting either indirect constraints from flavour physics \cite{Mahmoudi:2009zx}, or the direct LHC searches for charged Higgs bosons \cite{Aad:2012tj, *CMSHp}; again, since the charged scalar does not couple to fermions. On the other hand, values for $M_{\Hp} \sim 100\GeV$ would limit the mass for the IDM dark matter candidate, $M_{\rm{LOP}}$, to the low mass regime. 

As can be seen from Fig.~\ref{Fig:hgaga}, an $\Rgaga$ enhancement is not compatible with requiring a scenario valid up to the GUT scale, since these scenarios (black points) always lead to a suppressed value, $\Rgaga\leq 1$. The explanation behind this is given in the lower left plot, which shows $\Rgaga$ versus $\lambda_3$. Here it can be seen that, once the WMAP constraints are applied, the sign of $\lambda_3$ must be positive for scenarios to be viable up to the GUT scale. At the same time, the sign of this parameter governs the interference between the charged scalar contribution and the (SM) $WW$ contribution, and a suppression (enhancement) is obtained for positive (negative) values of $\lambda_3$.

\begin{figure}
	\begin{center}
		\includegraphics[width=0.95\columnwidth]{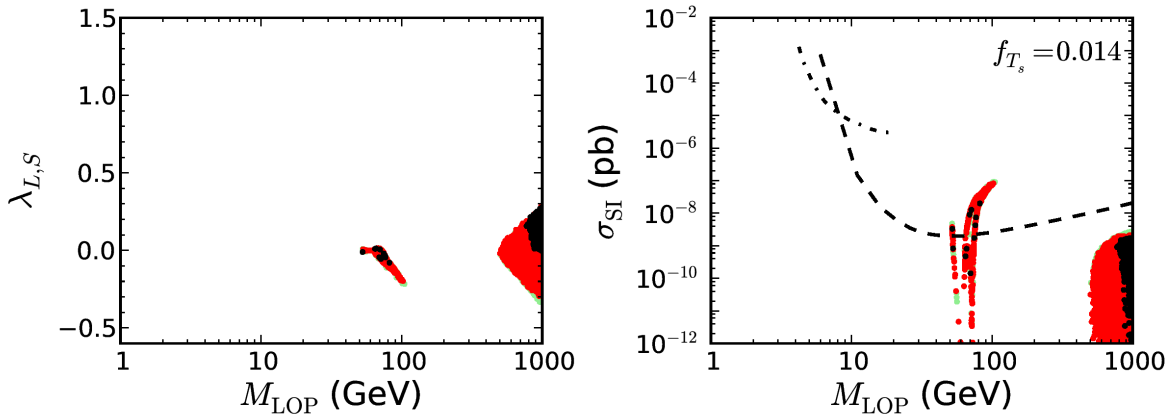}
	\end{center}
	\vspace{-2em}
	\caption{Projection of the viable IDM parameter space for $M_{\rm LOP}$ versus $\lambda_{L,S}$ (left) and $\sigma_{\rm SI}$ (right), with the limits from XENON10 (dash-dotted line) and XENON100 (dashed line). The colour coding is the same as in Fig.~\ref{Fig:RelicDensity1}. The full DM constraint from WMAP is applied, as well as the upper bound $\mathrm{BR}(\hn \to \mathrm{invisible})<0.65$.}
	\label{Fig:ATLASinv}
\end{figure}
The second set of plots in Fig.~\ref{Fig:hgaga} (right) are the same as those to the left, except that both the upper and the lower WMAP limits on the relic density have now been applied. These plots show important difference compared to the results where only the upper WMAP limit is applied. First of all, the number of allowed points is naturally much reduced. A second, very important, difference is the complete absence of points with $\Rgaga>1$ in this case. Once the (full) relic density constraint is imposed, the IDM rate for $\hn\to \gamma\gamma$ should therefore be expected to come out similar to (or below) that in the SM.

Another possibility, which is visible in Fig.~\ref{Fig:hgaga}, is to have more than an order of magnitude suppression of $\Rgaga$. This large suppression can be caused by ``invisible'' Higgs decays to the LOP, $\hn\to \Hn\Hn$ ($\hn\to \An \An$) for $\Mlop < M_{\hn}/2$. The resulting low values of $\Rgaga$ are not compatible with the observation of a SM-like Higgs boson at the LHC, which can serve as a further constraint on these scenarios. In addition, the ATLAS Collaboration has recently published upper limits \cite{ATLASinv} on the possible size of an invisible branching ratio for a SM-like Higgs boson. At the observed value for the Higgs mass, this corresponds to the constraint
\begin{equation}
	\mathrm{BR}(\hn \to \mathrm{invisible})<0.65
\label{eq:inv}
\end{equation}
at $95\%$ CL. In the IDM, when $\hn$ decays to LOP pairs is open, the invisible branching ratio tends to become very large. The effect of applying Eq.~\eqref{eq:inv} as a constraint can be seen in Fig.~\ref{Fig:ATLASinv}, which shows the valid points in the $(M_{\rm LOP},\lambda_{L,S})$ and $(M_{\rm LOP},\sigma_{\rm SI})$ planes after applying this bound. As illustrated by this figure, this constraint from the LHC provides a complementary limit to that from direct detection, in ruling out IDM dark matter in the low- and intermediate mass regimes.

\subsection{Benchmark scenarios}

As a final result of our analysis, we extract three benchmark points which capture different aspects of the IDM phenomenology. The scalar masses and quartic coupling for these points are summarized in Tab.~\ref{tab:bench}.

The first benchmark point has $\Mlop = 66\GeV \sim M_{\hn}/2$. It lies at the border between the Higgs-funnel and $WW^*$ threshold regions, where the annihilation proceeds partly through a $h^0$ resonance and partly into (virtual) gauge bosons. For the given point, the combination of the two mecanisms leads to the value of $\Omega_{\rm LOP}h^2 = 0.1113$, which is in agreement with the WMAP interval of Eq.~\eqref{Eq:RelicDensity}. 
The second benchmark scenario, point II, lies in the high-mass regime, the LOP being the pseudoscalar $\An$. Its mass is almost degenerate with the other inert scalars, so that the relic density of $\Omega_{\rm LOP}h^2 = 0.1224$ is achieved through the interplay of pair-annihilation and co-annihilation between $\An$, $\Hn$, and $\Hp$. Both scenarios I and II can be extrapolated to the GUT scale, but neither of them present an enhanced $\hn \to \gamma \gamma$ decay rate as compared to the Standard Model. Point II is out of the reach of LHC, but might be probed by the XENON1T experiment. 

\begin{table}
\begin{tabular}{c|cccccc|cccc}
\hline
 & ~~$M_{\hn}$~~ & ~~$M_{\Hn}$~~ & ~~$M_{\An}$~~ & ~~$M_{\Hp}$~~ & ~~$\lambda_L$ ~~& ~~$\lambda_2$~~ & ~~$\Omega_{\mathrm{LOP}}h^2$~~ & ~~$\sigma_{\mathrm{SI}}$ (pb)~~ & ~~$\Rgaga$~~ & ~~$\Lambda$ (GeV)~~\\
\hline 
I  & 125.3 & 66  & 131 & 137 & 0.012 & $9\cdot 10^{-4}$ & 0.1113 & $7.1\cdot 10^{-10}$ & 0.91 & $> 10^{16}$\\
II & 125.8 & 973 & 972 & 976 & 0.170 & 0.022 & 0.1224 & $4.4\cdot 10^{-10}$ & 1.00 & $> 10^{16}$ \\
III & 125.9 & 91 & 262 & 102 & -0.359 & 1.182 & $7\cdot 10^{-4}$ & $3\cdot 10^{-7}$ & 1.25 & $1.2\cdot 10^4$ \\
\hline
\end{tabular}
\caption{Scalar masses (in GeV) and quartic couplings at the input scale, $M_Z$, for three benchmark scenarios, with the corresponding values for the relic density, $\Omega_{\mathrm{LOP}}h^2$, direct detection cross section, $\sigma_{\mathrm{SI}}$, the $\hn\to \gamma\gamma$ rate relative to the SM, $\Rgaga$, and the maximum extrapolation scale, $\Lambda$.}
\label{tab:bench}
\end{table}

\begin{figure}
	\begin{center}
		\includegraphics[width=\columnwidth]{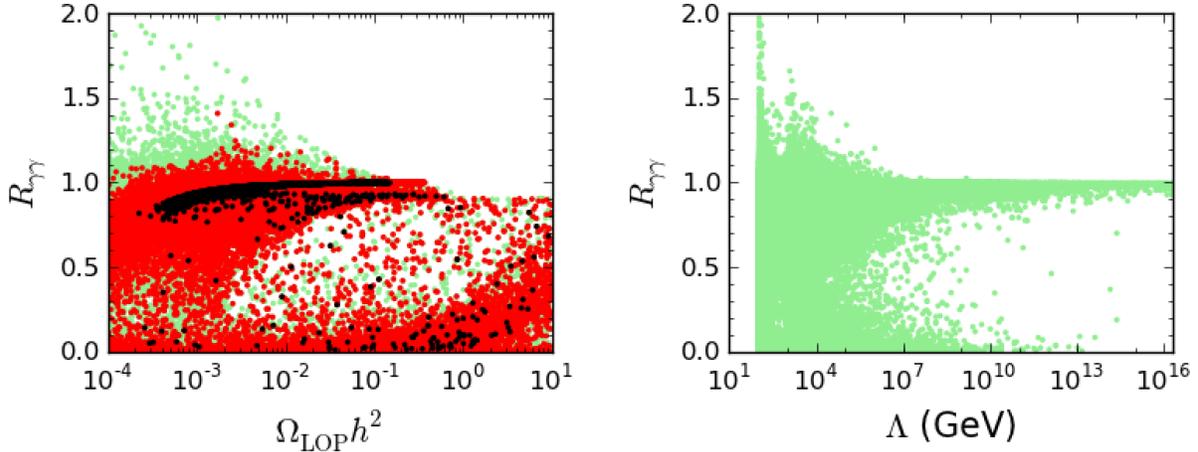}
	\end{center}
	\vspace{-2em}
	\caption{Correlation of the relative branching ratio $\Rgaga=\mathrm{BR}(\hn \rightarrow \gamma \gamma)_{\mathrm{IDM}}/\mathrm{BR}(\hn \rightarrow \gamma \gamma)_{\mathrm{SM}}$ and the dark matter relic density $\Omega_{\rm LOP}h^2$ (left) and the maximum extrapolation scale $\Lambda$ (right). The colour coding on the left-hand side is the same as in Fig.~\ref{Fig:RelicDensity1}.}
	\label{Fig:hgagacorr}
\end{figure}

Finally, point III is inspired by recent LHC observations rather than by its dark matter phenomenology. It presents an enhanced $\hn \to \gamma \gamma$ decay rate ($\Rgaga = 1.25$) as currently favoured by ATLAS \cite{ATLASgaga}, but not by CMS \cite{CMSgaga}. This scenario has a rather low relic density ($\Omega_{\rm LOP}h^2 \sim 7\cdot 10^{-4}$). Extrapolation of this scenario is only possible up to a maximum scale of about 10 TeV, where the requirement of perturbativity fails. Let us note that in principle it would have been possible to find a scenario with $\Rgaga\sim 1.25$ and a higher extrapolation scale (see e.g.~Fig.\ \ref{Fig:hgagacorr}), but the corresponding charged scalar masses are found to be very close to the limit $M_{\Hp} > M_{W}$. We have therefore chosen a more conservative point where $M_{\Hp} > 100\GeV$.

The fact that it is not possible to satisfy all the constraints which have been discussed in our detailed analysis while obtaining an enhanced $\Rgaga$ within the IDM is again illustrated in Fig.\ \ref{Fig:hgagacorr}. This figure shows the correlation of $\Rgaga$ to the relic density $\Omega_{\rm LOP}h^2$ (left) and the maximum extrapolation scale $\Lambda$ (right) obtained from our parameter space scan. We clearly see that points with $\Rgaga>1$ are only found for $\Omega_{\rm LOP}h^2 < 0.1$. In the high-mass regime, due to the requirement $M_{\Hp} > M_{\rm LOP}$, the charged scalar is also rather heavy, and the branching ratio of the decay $\hn \to \gamma \gamma$ is equal to the SM value. In the low-mass regime, the LOP is light enough to allow for invisible decay of the SM-like Higgs, so that its decay rate into photons is suppressed. In the intermediate mass region, as mentioned before, $\lambda_3$ is restricted to positive values in order to satisfy the upper and lower WMAP limits. At the same time, a negative value of $\lambda_3$ is necessary to achieve $\Rgaga>1$ (see Fig.\ \ref{Fig:hgaga}). Consequently, both conditions cannot be satisfied at the same time. Relaxing the lower WMAP bound allows for points similar to our benchmark scenario II, i.e.\ presenting an enhanced value of $\Rgaga>1$, but with a lower dark matter relic density $\Omega_{\rm LOP}h^2 \lesssim 0.1$. 

Similarly, the situation of $\Rgaga>1$ occurs only for extrapolation scales lower than $\Lambda \lesssim 10^7$ GeV, as can be seen from Fig.~\ref{Fig:hgagacorr} (right). As discussed before, the extrapolating to high scales puts an upper limit on the absolute value of the coupling parameters. In consequence, the decay rate of $\hn \to \gamma \gamma$ is smaller because of the limited quartic coupling $\lambda_3$.

\section{Summary \label{Sec:Discussion}}

We have presented an extensive study of the Inert Doublet Model in the light of the recent potential discovery of a Higgs-like boson with a mass around $126$ GeV at the Large Hadron Collider (LHC).  
To this end, we have computed the one-loop corrections to the scalar masses of the model and evaluated their numerical effect. In particular, we have shown the difference between the pole mass and the running \MSbar\ masses to be of the order of a few percent. This is sizeable in the context of dark matter annihilation, which often relies on precise mass differences with respect to a resonance, or on the presence of co-annihilation which is equally sensitive to the precise relations among the masses of the involved particles.

Moreover, we examined the high-energy behaviour of the IDM when treating the Higgs boson mass either as a free parameter, or under the assumption that the particle observed at the LHC corresponds to the IDM Higgs boson. In particular, we require the perturbativity of all couplings, stability of the electroweak vacuum, and unitarity of the $S$-matrix. By employing the one-loop renormalization group equations, we have shown that these constraints become less prohibitive than in the case of the Standard Model. 
More precisely, the IDM can be extrapolated up to high scales, including up to the GUT scale, for a non-negligible range of parameter choices and for larger ranges of the Higgs boson mass, with the experimentally observed value lying firmly within the region where the IDM can remain valid up to very high energy scales. 

We have then imposed a series of constraints coming from collider (LEP, oblique parameters) and dark matter (WMAP, direct detection) experiments, combining them with the aforementioned theoretical requirements. We find that the oblique parameter constraints has only mild effects upon the IDM parameter space, and the main result of the LEP bounds is to push the NLOP mass above $100$ GeV, which in turn results in the absence of neutal co-annihilation in the lower LOP mass regime (roughly below $80$ GeV). 

As a further consequence, the relic density in this mass region is driven by the self-annihilation cross section, which can be modified by tuning the couplings of the LOP ($\lambda_{L}$ or $\lambda_{S}$ for $\Hn$ or $\An$, respectively) to the Higgs boson, or by approaching the Higgs resonance and $WW$ production threshold. Far from the resonance/threshold, relatively large $\lambda_{L}$ (or $\lambda_{S}$) values are required for the WMAP bounds to be satisfied. This also implies a large WIMP-nucleon scattering cross section, which brings the model in tension with the current XENON bounds. The only surviving region lies close to the Higgs resonance/$WW$ production threshold, where $\lambda_{L,S}$, and consequently the coupling to quarks via Higgs exchange, has to be small in order for the depletion rate of DM not to be too high. 
The significant reduction of the available parameter space demonstrates the crucial character Higgs measurements can have for a number of dark matter models. Another measurement which we have used to constrain the mass region $\Mlop < M_{\hn}/2$ is the upper limit on an invisible branching ratio for the SM-like Higgs boson observed at the LHC. Here we find that the current limits are complementary, and give results that are very similar in reach, to those from direct detection experiments.

The high mass regime, on the other hand, remains relatively unaffected by all the current bounds, and we find a viable region of parameter space for DM masses above $500\GeV$. This provides an interesting scenario for direct detections experiments, since it was recently pointed out \cite{Klasen2013EWDD} that the future XENON-1T experiment should be able to probe essentially the entire viable IDM parameter space, including the high-mass regions that are likely to remain inaccessible to the LHC. 

Finally, we have studied the interplay between the dark matter phenomenology and the LHC Higgs phenomenology, in particular the possibility to have a modified $\hn \to \gamma\gamma$ decay rate, as still allowed by the present data. Here, the most important result of our analysis is that enhanced $\hn \to \gamma\gamma$ rate can only be achieved for a dark matter relic density which is below the WMAP limit. Moreover, we find that the corresponding parameter points cannot be extrapolated to scales higher than $10^7$ GeV with the requirement of perturbativity satisfied. Based on our findings, we have presented three benchmark scenarios that capture the essential features discussed throughout our analysis, i.e.\ dark matter relic density and detection, possibility to extrapolate the model to high scales, and an enhanced $\hn \to \gamma\gamma$ decay rate inspired by the LHC discovery and the current excess observed in this channel by ATLAS.

\acknowledgments

The authors would like to thank S.~Rydbeck for valuable discussions. We are grateful to F.~Staub and B.~Fuks for their help concerning 
the use of \SARAH\ and \FeynRules. Finally, we would like to thank G.~B\'elanger and A.~Pukhov for substantial help with the \MO\ code and 
especially for providing us with a version computing three-body final state contributions. The work of A.G.\ and B.H.\ was in part supported 
by the Landesexzellenz-Initiative Hamburg. During part of this work O.S.\ was supported by the Collaborative Research Center SFB 676 of the 
DFG, ``Particles, Strings, and the Early Universe''. O.S.\ is now supported by the Swedish Research Council (VR) through the Oskar Klein 
Centre. At LAPTh, this activity was developed coherently with the research axes supported by the LABEX grant ENIGMASS. 
The authors would like to thank the DESY Theory Group, where part of this work was performed, for warm hospitality.
\newpage

\appendix
\section{One-loop scalar masses \label{App:ScalarMasses}}
\vspace{-0.5em}

\allowdisplaybreaks
In this Appendix, we give the expressions for the (one-loop) running scalar masses in the Inert Doublet Model, which have been obtained as discussed in Sec.\ \ref{Sec:IDMoneloop}. In the \MSbar\ scheme, the masses for the SM-like Higgs boson ($\hn$) and inert scalars ($\Hn$, $\An$, $\Hp$) are given by
\begin{equation}
\begin{aligned}
		\lm_{\hn}^2 = &m_{\hn}^2-\frac{\alpha}{16 \pi  c_W^2 s_W^2}\Biggl\{
		c_W^2\left(6+\frac{m_{\hn}^2}{M_W^2}\right)A_0(M_W^2)
		+\left(3+\frac{m_{\hn}^2}{2M_Z^2}\right)A_0(M_Z^2)\\
	&+c_W^2M_W^2\left(12+\frac{m_{\hn}^4}{M_W^4}\right)B_0(M_W^2,M_W^2)+M_Z^2\left(6+\frac{m_{\hn}^4}{2M_Z^4}\right)B_0(M_Z^2,M_Z^2)\\
	&+\frac{3m_{\hn}^2}{2M_Z^2}\Bigl[A_0(m_{\hn}^2)+3m_{\hn}^2B_0(m_{\hn}^2,m_{\hn}^2)\Bigr]-6M_Z^2-12c_W^2M_W^2\Biggr\}\\
	&+\frac{1}{8\pi^2}\sum_f N_C^f y_f^2\Bigl[A_0(m_f^2)+y_f^2v^2B_0(m_f^2,m_f^2)\Bigr]\\	
	& -\frac{1}{16\pi^2}\Biggl\{\lambda _L A_0(m_{\Hn}^2)+\lambda_S A_0(m_{\An}^2) +\lambda _3 A_0(m_{\Hp}^2)\\
	&+2 \lambda _L^2 v^2 B_0(m_{\Hn}^2,m_{\Hn}^2)+2\lambda _S^2v^2  B_0(m_{\An}^2,m_{\An}^2)+\lambda _3^2v^2  B_0(m_{\Hp}^2,m_{\Hp}^2)\Biggr\},
\end{aligned}
\end{equation}
\vspace{-0.5em}
\begin{equation}
\begin{aligned}
		\lm_{\Hn}^2 = &m_{\Hn}^2-\frac{\alpha}{16\pi c_W^2 s_W^2} \Biggl\{3\Bigl[A_0(M_Z^2)+2c_W^2A_0(M_W^2)\Bigr]+2c_W^2m_{\Hp}^2B_0(m_{\Hp}^2,M_W^2)\\
		&+m_{\An}^2B_0(m_{\An}^2,M_Z^2)-2M_Z^2-4c_W^2M_W^2\Biggr\}\\
		& -\frac{1}{16\pi^2} \Biggl\{\lambda_3 A_0(M_W^2)+\lambda_S A_0(M_Z^2)+\lambda_L A_0(m_{\hn}^2)+3\lambda_2A_0(m_{\Hn}^2)\\
		& +\lambda_2A_0(m_{\An}^2)+2\lambda_2A_0(m_{\Hp}^2) +\frac{1}{2}(\lambda_4+\lambda_5)^2v^2B_0(m_{\Hp}^2,M_W^2)\\
		&+\lambda_5^2 v^2B_0(m_{\An}^2,M_Z^2)+4\lambda_L^2 v^2 B_0(m_{\Hn}^2,m_{\hn}^2)\Biggr\},
\end{aligned}
\end{equation}
\vspace{-1em}
\begin{equation}
\begin{aligned}
		\lm_{\An}^2 &= m_{\An}^2-\frac{\alpha}{16\pi c_W^2 s_W^2} \Biggl\{3\Bigl[A_0(M_Z^2)+2c_W^2A_0(M_W^2)\Bigr]+2c_W^2m_{\Hp}^2B_0(m_{\Hp}^2,M_W^2)\\
		&+m_{\Hn}^2B_0(m_{\Hn}^2,M_Z^2)-2M_Z^2-4c_W^2M_W^2\Biggr\}\\
		&-\frac{1}{16\pi^2} \Biggl\{\lambda_3 A_0(M_W^2)+\lambda_L A_0(M_Z^2)+\lambda_S A_0(m_{\hn}^2)+\lambda_2A_0(m_{\Hn}^2)\\
		&+3\lambda_2A_0(m_{\An}^2)+2\lambda_2A_0(m_{\Hp}^2) +\frac{1}{2}(\lambda_4-\lambda_5)^2v^2B_0(m_{\Hp}^2,M_W^2)\\
		&+\lambda_5^2 v^2B_0(m_{\Hn}^2,M_Z^2)+4\lambda_S^2 v^2 B_0(m_{\An}^2,m_{\hn}^2)\Biggr\},
\end{aligned}
\end{equation}
\begin{equation}
\begin{aligned}
		\lm_{\Hp}^2 &= m_{\Hp}^2-\frac{\alpha}{16\pi c_W^2 s_W^2} \Biggl\{6c_W^2A_0(M_W^2)+3(c_W^2-s_W^2)^2A_0(M_Z^2)\\
		&+(c_W^2-s_W^2)^2m_{\Hp}^2B_0(m_{\Hp}^2,M_Z^2)+c_W^2m_{\Hn}^2B_0(m_{\Hn}^2,M_W^2)\\
	&+c_W^2m_{\An}^2B_0(m_{\An}^2,M_W^2)-2M_Z^2(c_W^2-s_W^2)^2-4c_W^2M_W^2 \Biggr\}	\\
		&-\frac{1}{16\pi^2} \Biggl\{ \left(\lambda_3+\lambda_4\right) A_0(M_W^2)+\frac{\lambda_3}{2} A_0(M_Z^2)+\frac{\lambda_3}{2} A_0(m_{\hn}^2)+\lambda_2A_0(m_{\Hn}^2)\\
	&+\lambda_2A_0(m_{\An}^2)+4\lambda_2A_0(m_{\Hp}^2)+\lambda_3^2v^2B_0(m_{\hn}^2,m_{\Hp}^2)	\\
	&+\frac{1}{4}\Bigl[(\lambda_4+\lambda_5)^2v^2B_0(m_{\Hn}^2,M_W^2)+(\lambda_4-\lambda_5)^2v^2B_0(m_{\An}^2,M_W^2)\Bigr]\Biggr\}.
\end{aligned}
\end{equation}
In these expressions $M_Z$ is the $Z$ boson mass, $M_W$ the $W$ mass, and $y_f$ ($m_f$) the Yukawa coupling (mass) of a generic fermion $f$ that comes in $N_C^f$ number of colours ($3$ for quarks and $1$ for leptons). We further denote the electromagnetic fine-structure constant as $\alpha$, and we use the shorthand notation $c_W^2=1-s_W^2=M_W^2/M_Z^2$ for the weak mixing angle. The functions $A_0$ and $B_0$ are the regularized (finite) real parts of the one- and two-point scalar integrals, respectively. The first argument $p^2=0$ of the function $B_0$ has been suppressed. For details on these integrals, as well as their expressions in terms of elementary functions, the reader is referred to Ref.\ \cite{Denner:1991kt}. 


\section{One-loop beta functions \label{App:BetaFunctions}}

The one-loop beta functions for the running of the parameters appearing in the general 2HDM potential are known 
in the literature, see for example \cite{Ferreira:2009jb}. To collect everything using a consistent notation for 
the IDM, we give the results of our calculation here. The beta function for a running parameter $\lambda_{i}(Q)$ 
is defined through
\begin{equation}
	16\pi^2 \frac{\partial\lambda_{i}}{\partial \log Q} = 
	\beta_{\lambda_{i}} = \beta^{(s)}_{\lambda_{i}}+\beta^{(g)}_{\lambda_{i}}+\beta^{(y)}_{\lambda_{i}},
\end{equation}
where we choose to separate the contributions from scalars $(s)$, gauge interactions $(g)$ and Yukawa couplings $(y)$. 
The contributions to the IDM beta functions from the scalar interactions are given by
\begin{align}
	\beta_{\lambda_1}^{(s)} &= 24 \lambda_1^2 + 2 \lambda_3^2 + 2 \lambda_3 \lambda_4 +  \lambda_4^2 +  \lambda_5^2, \\
	\beta_{\lambda_2}^{(s)} &= 24 \lambda_2^2 + 2 \lambda_3^2 + 2 \lambda_3 \lambda_4 +  \lambda_4^2 +  \lambda_5^2, \\
	\beta_{\lambda_3}^{(s)} &= 4\left( \lambda_1 + \lambda_2 \right) \left( 3 \lambda_3 +  \lambda_4 \right) + 4 \lambda_3^2 + 2 \lambda_4^2 + 2 \lambda_5^2, \\
   \beta_{\lambda_4}^{(s)} &= 4 \lambda_4 \left( \lambda_1 + \lambda_2 + 2 \lambda_3 +  \lambda_4 \right) + 8 \lambda_5^2, \\
   \beta_{\lambda_5}^{(s)} &= 4 \lambda_5 \left( \lambda_1 + \lambda_2 + 2 \lambda_3 + 3 \lambda_4 \right).
\end{align}
\newpage
The contributions from gauge interactions read
\begin{eqnarray}
	 \beta_{\lambda_1}^{(g)} &=&  \frac{3}{8} \left( 3 g^4 + g'^4 + 2 g^2 g'^2 \right) - 3 \lambda_1 \left( 3 g^2 + g'^2 \right), \\
	\beta_{\lambda_2}^{(g)} &=& \frac{3}{8} \left( 3 g^4 + g'^4 + 2 g^2 g'^2 \right) - 3 \lambda_2 \left( 3 g^2 + g'^2 \right), \\
 \beta_{\lambda_3}^{(g)} &=& \frac{3}{4} \left( 3 g^4 + g'^4 - 2 g^2 g'^2 \right) - 3 \lambda_3 \left( 3 g^2 + g'^2 \right), \\
\beta_{\lambda_4}^{(g)} &=& 3 g^2 g'^2 - 3 {\lambda_4} \left( 3 g^2 + g'^2 \right) , \\
	 \beta_{\lambda_5}^{(g)} &=&  - 3 \lambda_5 \left( 3 g^2 + g'^2 \right).
\end{eqnarray}
Finally, including only the third generation fermions, we get from the Yukawa interactions
\begin{eqnarray}
	\beta_{\lambda_1}^{(y)} &=&  4 \lambda_1 \left( y_\tau^2 + 3 y_b^2 + 3 y_t^2 \right) - 2 \left( y_\tau^4 + 3 y_b^4 + 3 y_t^4 \right), \\
	\beta_{\lambda_2}^{(y)} &=& 0 ,\\
	\beta_{\lambda_3}^{(y)} &=& 2 \lambda_3 \left( y_\tau^2 + 3 y_t^2 + 3 y_b^2 \right) , \\
	\beta_{\lambda_4}^{(y)} &=& 2 \lambda_4 \left( y_\tau^2 + 3 y_t^2 + 3 y_b^2 \right) , \\
	\beta_{\lambda_5}^{(y)} &=& 2 \lambda_5 \left( y_\tau^2 + 3 y_t^2 + 3 y_b^2 \right).
\end{eqnarray}
We also need the running of the gauge couplings, which is given by 
\begin{equation}
	 \beta_{g_i} = b_i g_i^3 \quad (g_i=g,g',g_s),
\end{equation}
where, for the case of two Higgs doublets, $b_1=-3$, $b_2=7$, and $b_3=-7$. The corresponding expressions for the Yukawa couplings are 
\begin{eqnarray}
	\beta_{y_t}^{(y)} &=&  y_t \left( -\frac{17}{12} g'^2 - \frac{9}{4} g^2 - 8 g_s^2 + y_\tau^2 + \frac{9}{2} y_t^2  + \frac{3}{2} y_b^2 \right), \\
	\beta_{y_b}^{(y)} &=&  y_b \left( -\frac{5}{12} g'^2 - \frac{9}{4} g^2 - 8 g_s^2 + y_\tau^2  + \frac{3}{2} y_t^2 + \frac{9}{2} y_b^2\right), \\
	\beta_{y_\tau}^{(y)} &=&  y_\tau \left( -\frac{15}{4} g'^2 - \frac{9}{4} g^2 + \frac{5}{2} y_\tau^2 + 3 y_t^2 + 3 y_b^2 \right).
\end{eqnarray}

\bibliographystyle{JHEP}
\bibliography{IDM}
\end{document}